# A determination of $F_L$ at $x = Q^2/s$ with HERA data

Frank E. Taylor

Department of Physics
Laboratory for Nuclear Science
Massachusetts Institute of Technology
Cambridge, MA 02139

June 26, 2024

## ABSTRACT

It is well known that there are persistent statistical tensions with the standard model in the low $Q^2$ HERA deep inelastic scattering neutral current data characterized by a turn-over of $F_2(x, Q^2)$ at low x and low $Q^2$. One important experimental signature that sheds light on this low $Q^2$ region is the determination of the longitudinal structure function $F_L(x, Q^2)$. This paper describes a novel method to determine $F_L$ based on an extrapolation of the reduced NC cross section at fixed s and Q to the minimum value of x given by $Q^2/s$. At this kinematic point, the reduced cross section equals $2xF_1 = F_2 - F_L$ so that a determination of both this value and the value of $F_2$, determines $F_L$. Since the polarization of the exchanged photon is transverse at this kinematic point, we expect $F_L$ to be small because its dominate gluon component is strongly suppressed. Surprisingly, we find $F_L$ at low $Q^2$ to be much larger than expectation and observe that both $F_L$ and $F_2$ at $x = Q^2/s$ show several properties consistent with the dipole picture. We discuss the statistical as well as chief systematic errors of our method and we tabulate our determinations of $F_2$, $2xF_1$ and $F_L$ in the Appendix.

*Corresponding email: fet@mit.edu*

# I. INTRODUCTION

Probing the parton structure of the nucleon remains an enduring and sometimes obdurate quest for both theoretical as well as experimental high energy physicsts. From the landmark measurements made by the SLAC-MIT group at SLAC and the theoretical predictions of Bjorken and physical interpretations of Feynman, studying the physics of lepton-nucleon scattering remains productive. The experimental program at HERA-DESY in Hamburg [1] extended the kinematic range of experimental data taken earlier at SLAC and CERN and FNAL to very low x and $Q^2$ as well as high s and $Q^2$. It is in this kinematic low region where putative quark threshold effects, quark masses and longitudinal polarized virtual Compton scattering become operative. In the early 2030s, the Electron-Ion Collider at BNL [2] will contribute data of very high statistical value in not only e-p scattering but also e-A scattering, where A is a heavy nucleus. However, data from the EIC program will be at lower s ~2 x $10^4$ $GeV^2$ vs. the flagship HERA dataset where s ~1 x $10^5$ $GeV^2$.

Many analyses of the HERA NC DIS (Neutral Current Deep Inelastic Scattering) inclusive data discuss the statistical tension with the standard model DGLAP (Dokshitzer–Gribov–Lipatov–Altarelli–Parisi) analysis in the low x, low $Q^2$ HERA NC data. Global fits have been performed to NLO (Next to Leading Order) and NNLO (Next to Next to Leading Order) models that result in persistently poor results. These fits are described in detail in the definitive combined H1 and ZEUS analysis of Abramowicz et al. [3] [4] where it is noted that typically the $\chi^2$/d.o.f = 1.15 to 1.2 for > 1000 d.o.f. (degrees of freedom). For example, the NLO fit yields $\chi^2$/d.o.f = 1.2 for 1131 d.o.f. (Table 4 of [4]). This fit has a p value = 3.6 x $10^{-6}$. And much to the consternation of these authors, the $\chi^2$/d.o.f. is not improved by a NNLO calculation. The authors point to two troublesome regions where the tension with the SM (Standard Model) arise. The region $3.5 < Q^2 < 15$ $(GeV/c)^2$ gives roughly 1/3 of the $\chi^2$ to the SM QCD (Standard Model Quantum Chromo Dynamics) fits, whereas the region $Q^2 > 150$ $(GeV/c)^2$ contributes the remaining 2/3. And there is a lot of tension at the very low $Q^2$ region < 4.5 $(GeV/c)^2$, where it is apparent that the reduced cross section turns over at the low x corner [5]. Further, some of the SM fits suggest that in the very low x - low $Q^2$ region the gluon distribution is suppressed which is contra to expectations [3].

The reduced cross section for inclusive e-p DIS tabulated in the HERA compilation [4] is given by:

$$\sigma_r\left(x,Q^2,s\right)=\frac{d^2\sigma(e^{\pm}p)}{dQ^2dx}\left[\frac{Q^4}{2\pi\alpha_e^2}\frac{x}{Y_+}\right]=F_2-\frac{y^2}{Y_+}F_L=2xF_1+\frac{2(1-y)}{Y_+}F_L, \qquad (1)$$

where the usual DIS kinematic variables in Eq. 1 are defined as below:

$$x=\frac{Q^2}{4E_eE_p\left[1-\frac{E_e^{'}}{2E_e}(1+\cos\theta)\right]}=\frac{Q^2}{sy}, \qquad (2)$$

and $E_e$ ($E_p$) are the energies of the incident electron (proton), respectively, $s = 4E_eE_p$ is the total energy squared of the e-p system, $\theta$ is the polar angle of the scattered electron, $Y_+ = 1 + (1 - y)^2$ and the inelasticity variable y is defined by Eq. 2. The structure function $2xF_1$ corresponds to transversely polarized virtual photon scattering off spin ½ partons. In the limit of no longitudinal scattering term, the Callen-Gross relation applies making $F_2 = 2xF_1$.

This tension is illustrated in Fig. 1 taken from [6]. Many explanations of this low x, low Q turn-over behavior have been suggested. These include the hypothesis that NNLO and NNLO+NNLx perturbative calculations involving resummations are needed [7], [8].

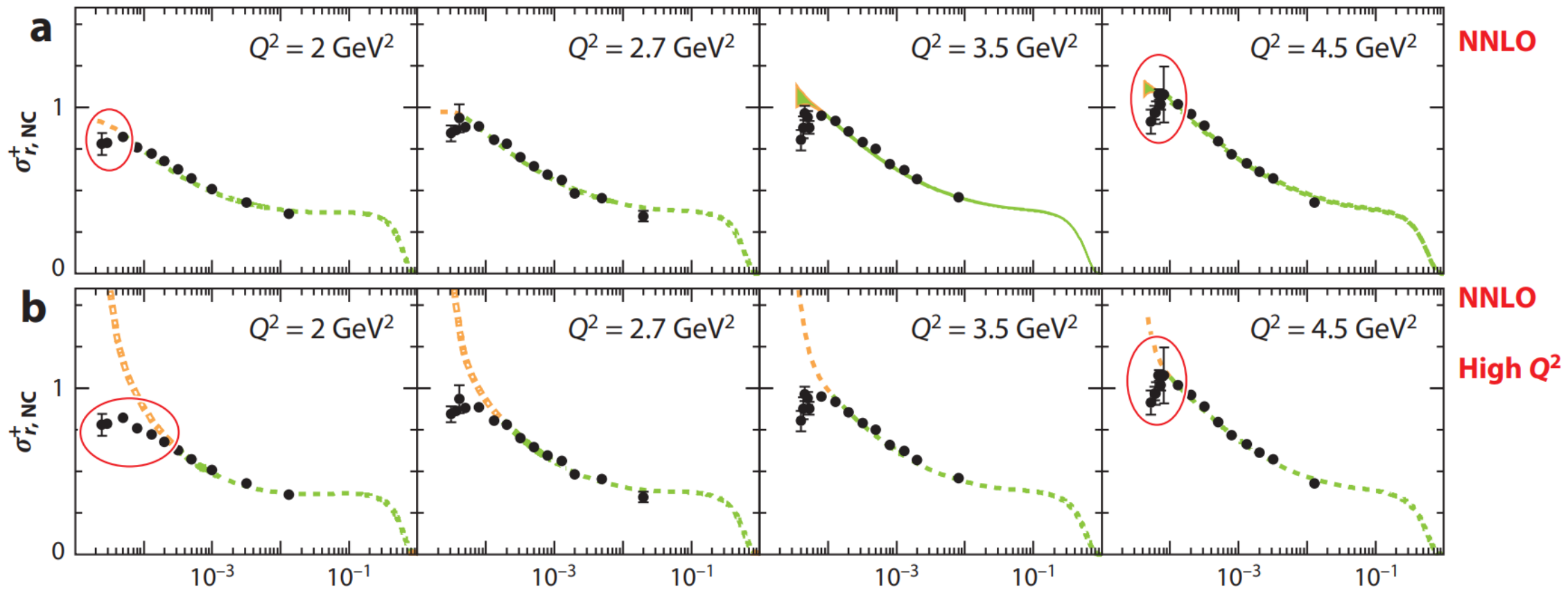

FIG.1: Shown are the values of the reduced cross section for low $Q^2$ as a function of x of the combined HERA H1-ZEUS from [6]. The turn-over at low x and low $Q^2$ is quite evident. It is this feature of the data that gives rise to some of the tension with the SM and is the subject of this paper.

Others have suggested that there are higher twist terms operative [9] that have the form $x \rightarrow x + D/Q^2$ and therefore affect the very low $Q^2$ physics. In fact, the very question of whether or not perturbative methods really do apply is a relevant one [10].

All of these considerations suggest that a study of the reduced cross section at the minimum value of $x_{min} = Q^2/s$, corresponding to $y = 1$, is interesting. However, this kinematic point is reached only when the probing electron backscatters through $2\pi$ and thus is inaccessible to direct measurement. Further, the region near this kinematic point involves difficult measurements of the scattered electron at low laboratory energies in large backgrounds. Despite these difficulties with an extrapolation procedure and a critical assumption of the behavior of the parton PDFs at this kinematic point, to be described in the next section, the reduced cross section at $x_{min}$ can be determined. That is the subject of this paper.

Crucial to understanding the physics of low x is the behavior of the electron – parton scattering when the polarization of the virtual photon can have both a transverse as well as longitudinal polarization depending on the y value. The separation of the cross sections for these two polarization states was worked out by Hand [11]. In the limit of neglecting masses, the longitudinal polarization of the virtual photon is given by:

$$\varepsilon = \left[1 + 2\left(\frac{\upsilon^2 + Q^2}{Q^2}\right)\tan^2\left(\frac{\theta}{2}\right)\right]^{-1} \approx \frac{2(1-y)}{1+(1-y)^2}, \tag{3}$$

where ν is the energy of the recoil hadron system, θ is the angle of the scattered electron and $Q^2$ and y are usual deep inelastic electron-proton variables defined by Eqs. 1 and 2. We note that at y = 1 the longitudinal polarization given by Eq. 3 is zero. Hence, a measurement of the reduced cross section at $x_{min} = Q^2/s$ is a determination of the transversely polarized photon – parton scattering.

In the usual case of the virtual photon scattering off spin ½ partons, the polarization of the photon has to be transverse. But in the longitudinal regimen the participation of quark-antiquark pairs and gluons in the proton distorts the very low x shape of the proton structure function. The magnitude of this distortion is directly dependent on the magnitude of the strong coupling constant. In terms of the scattered virtual photon of virtuality, - $Q^2$, the two processes are $\gamma^* + \left(q_i, \overline{q_i}\right) \rightarrow g + \left(q_f, \overline{q_f}\right)$ and $\gamma^* + g \rightarrow q + \overline{q}$ . These effects are captured by the expression of the longitudinal structure function $F_L$ based on the quark-parton model [12] given by:

$$F_L(x,Q^2) = \frac{\alpha_s(Q^2)}{\pi}\left[\frac{4}{3}\int_x^1 \frac{dy}{y}\left(\frac{x}{y}\right)^2 F_2(y,Q^2) + 2\sum e_q^2 \int_x^1 \frac{dy}{y}\left(\frac{x}{y}\right)^2\left(1-\frac{x}{y}\right) yG(y,Q^2)\right] \tag{4}$$

where $\alpha_s$ is the strong coupling constant, $F_2(x,Q^2)$ is the structure function of the virtual photon coupling to the charge of the participating quarks and antiquarks of the proton and $xG(x,Q^2)$ is the gluon distribution that can interact with the exchanged photon through its own coupling to quark-antiquark pairs and $\sum e_q^2$ is the sum of the charge-squared of the participating quark and antiquarks. Calculation to higher order changes $F_L$ of Eq. 4 by only a small amount [13]. In this work $\alpha_s(Q^2)$ is evaluated by a fit of the Particle Data Group (PDG) over the range $1.44 < Q < 3.9 \times 10^3$ GeV/c [14].

Most determinations of $F_L$ are focused on extraction of the gluon content of the nucleon that is expected to dominate. In fact, in DIS studies the measurement of $F_L$ is the most direct way to gain access to the gluon content of the proton. Not only do DIS analyses seem to confirm this [15], [16] but also electroproduction studies, such as in Korover and Milner [17], conclude that the gluon component in $F_L$ dominates.

Unlike the usual Rosenbluth analyses [18], where data taken at different s-values allow the y-dependence of the reduced cross section to be measured for fixed values of x and $Q^2$ thereby permitting a separation of $F_2$ from $F_L$, our determination of $F_L$ is based on measurements of the structure function difference, $2xF_1 = F_2 - F_L$, and $F_2$ along a specified kinematic line defined by $x_{min} = Q^2/s$. Taking the difference of these two measurements determines $F_L$ at $y = 1$. Following this line to the limit $y = 1$, the longitudinal polarization of the virtual photon becomes zero. Therefore, we would expect the dominant gluon component in our determination to be suppressed, resulting in $F_L$ at $x_{min}$ to be small. But other effects could be important at $x_{min}$ – such as parton saturation [19].

Many have pondered the physics of ep NC scattering at low x. One model considered is the dipole picture [20, 21, 22, 23] where the virtual photon fluctuates into a quark-antiquark pair, much like in vector meson dominance in photoproduction for real photons, which subsequently interacts with the proton. The proponents claim that this approach has broad applicability, such as describing DIS, exclusive vector meson production and inclusive diffractive processes [24]. Analyzing DIS at low x in this picture results in a determination of the gluon distribution of the proton relative to a simple form of the dipole nucleon scattering cross section.

Regardless of the underlying physical principles, this paper describes a determination of $F_2$ and $2xF_1$ by means of the HERA DIS reduced cross sections, thereby measuring $F_L = F_2 - 2xF_1$ at y =1 where the polarization of the virtual exchanged photon is transverse. Naively, we would expect that $F_L \rightarrow 0$ at this kinematic point. But we find that it remains sizable. Thus, we interpret the " $F_L$" term to be a correction to $F_2$ needed for it to satisfy the Callen-Gross relation.

## II. METHOD

The extraction of $F_L$ from the HERA neutral current ep DIS data is notoriously difficult. In what has become the standard analysis (Rosenbluth [18]), the ep cross sections are measured at various x and $Q^2$ values for various s values. This enables the y-dependence of the reduced cross section to be measured for the same x and $Q^2$ values thereby determining $F_2$ as well as $F_L$. In this manner both the x-dependence as well as the $Q^2$ dependence of $F_L$ are determined. Many such analyses have been performed such as those carried out by the H1 [15] and ZEUS [16] Collaborations.

Our method is different and is more direct, but is highly specific. We notice that at $x_{min} = Q^2/s$ that y = 1 and $Y_+ = 1 + (1 - y)^2 = 1$, so that the reduced cross section becomes simply the difference between $F_2$ and $F_L$ ($2xF_1 = F_2 - F_L$) when $Q^2 << M_z^2$. At this kinematic point the cross section is given by:

$$\lim_{y\to 1}\left[\sigma_r\left(x,Q^2,s\right)\right]=F_2(Q^2/s,Q^2)-F_L(Q^2/s,Q^2)=2xF_1\left(Q^2/s,Q^2\right), \qquad (5)$$

where we have ignored the very small nonsinglet term, $xF_3$. Each analysis for a given data set at fixed Q and s rests on separately determining $F_2$ and $2xF_1 = F_2 - F_L$, both evaluated at the y = 1 limit. This is accomplished by studying the behavior of the reduced cross section for two y cuts, one for the determination of $F_2$ and the other for $2xF_1$. Taking the difference of these two observables at the $x_{min}$ (y = 1) limit results in a determination of $F_L$ along the $x_{min}$ - $Q^2$ line. Thus, the systematic errors associated with different s-related luminosity errors do not contribute – an advantage. However, by taking this limit for each value of $Q^2$-s results only in measurement a given $x_{min}$ - $Q^2$ contour of $F_L$ rather than separately measuring both the x and $Q^2$ dependences accessible by the Rosenbluth method. In order to separate the x and $Q^2$ dependence one would still have to compare different s-dependent x – $Q^2$ contours. It is with irony that our method of determining the longitudinal structure function, $F_L$, is at the kinematic point where the longitudinal polarization of the virtual photon is zero.

In order to determine $F_2$ and $F_L$ at $x_{min}$, we perform minimum $\chi^2$ fits to the reduced cross sections of the combined H1-ZEUS HERA data [4] by a polynomial in the double natural logarithm given by ln(ln(1/x)) of the form:

$$\ln\left[F_2-\frac{y^2}{Y_+}F_L\right]=P_n\left[\ln\left(\ln\left(1/x\right)\right)\right], \qquad (6)$$

where x is the Bjorken scaling variable. The polynomial is typically a quadratic or a cubic (n = 2 or 3, respectively) in the double log variable. The resulting parameters of the fit enable the reduced cross section to be evaluated at $x_{min}$ under the key assumption that the structure functions are continuous out to the limit y = 1. In order to separate $F_2$ from $F_L$ two fits are performed (with systematic error corrections to be described later) – one for $y < y_{cut}$ extrapolated to $x_{min}$ to determine $F_2$ and the other for $y > y_{cut}$ extrapolated to determine the other observable $2xF_1$. By separately determining $F_2$ and $2xF_1$, $F_L$ itself is estimated by the difference $F_2 - 2xF_1$ at $x = Q^2/s$. The systematic corrections of the method are made by fitting a model of $F_2$ and $2xF_1$ from parameterizations of the quark and gluon distributions (CTEQ [25]), [26] by comparing the true values of $F_2$ and $2xF_1$ with their fitted value. Since the systematic corrections are model-against-model calculations, they are not strongly dependent on the specifics of the model.

Figure 2 illustrates the method for determining $F_2$ and $2xF_1$ for a typical value of $Q^2$ taken from Table 1 of the HERA data at $\sqrt{s} = 318$ GeV [4]. From the figure we notice that the $F_2$ determination tends to underestimate $F_2$, whereas the $2xF_1$ determination tends to overestimate its value at $x_{min}$. The differences of these estimates versus their true values are used to correct the corresponding fits to the data. This calculation is illustrated in the figure by the differences between the dotted (estimated) and solid (true) lines.

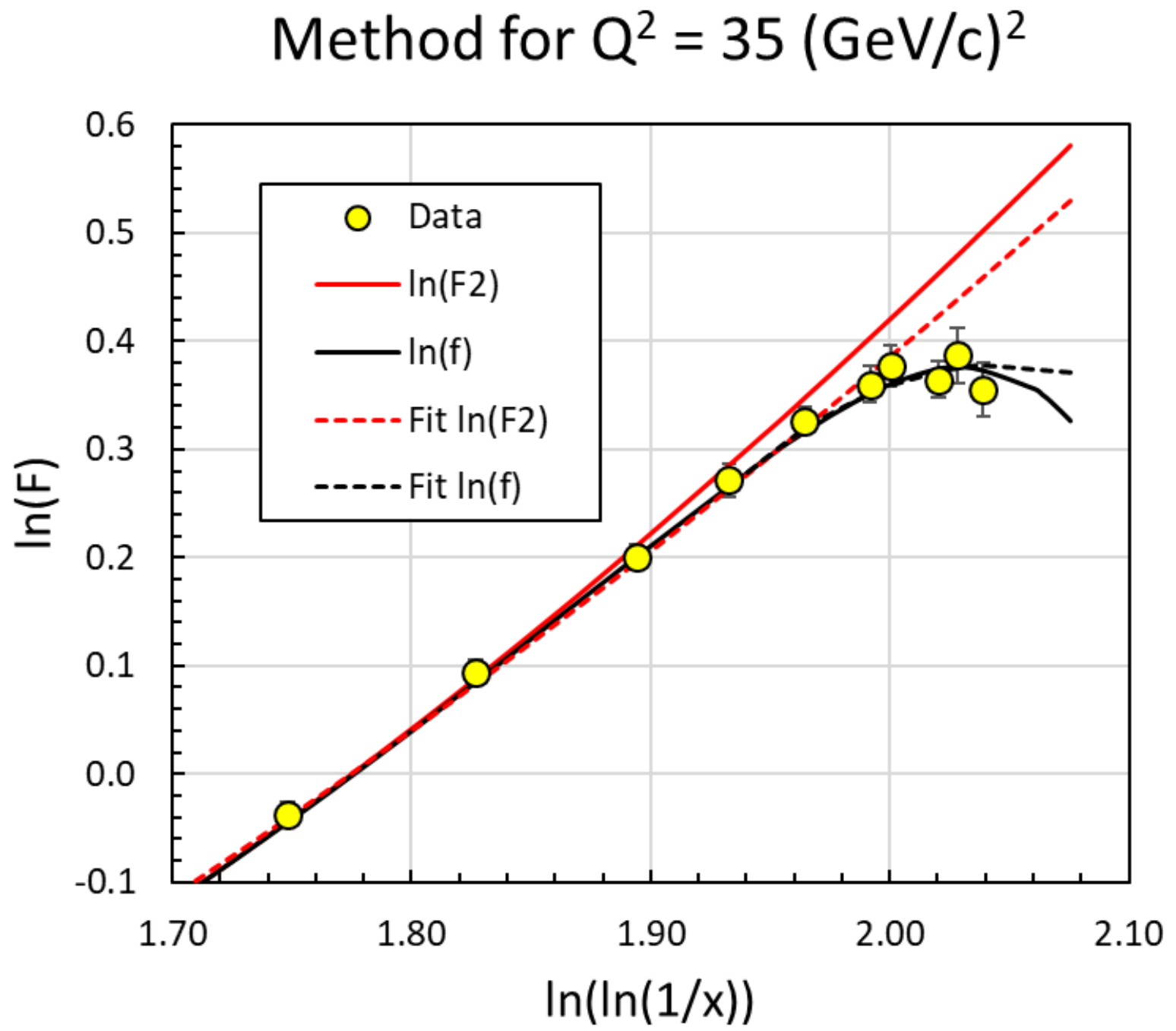

FIG. 2: Shown as yellow circles are the natural logs of the reduced cross section for $Q^2 = 35$ (GeV/c)$^2$ data as a function of the double log variable, ln(ln(1/x)). The CTEQ estimate of ln($F_2$) and of ln($xF_1$)=ln($F_2$ - ($y^2/Y_+$)$F_L$) are indicated as the solid lines. Two fits to the modeled data corresponding to the two y cuts are shown.

This fitting procedure is a large source of systematic errors. The polynomial terms are highly correlated so the error of the extrapolated $x_{min}$ point has to be estimated by the full covariance matrix of the fit. Typical $\chi^2$/d.o.f of the fitting procedure are shown in Fig. 3 for a three-parameter fit to the reduced cross section of Table 1 ($\sqrt{s}$ = 318 GeV) [4] to determine $F_2$ and $2xF_1$. The regions of the data used in the fits are $y < 0.3$ and $x < 0.33$ for $F_2$ and $y > 0.3$ for $2xF_1$. Four parameter fits were also performed where the data were extensive enough to permit them. The difference between the two fitting procedures was taken as a systematic error.

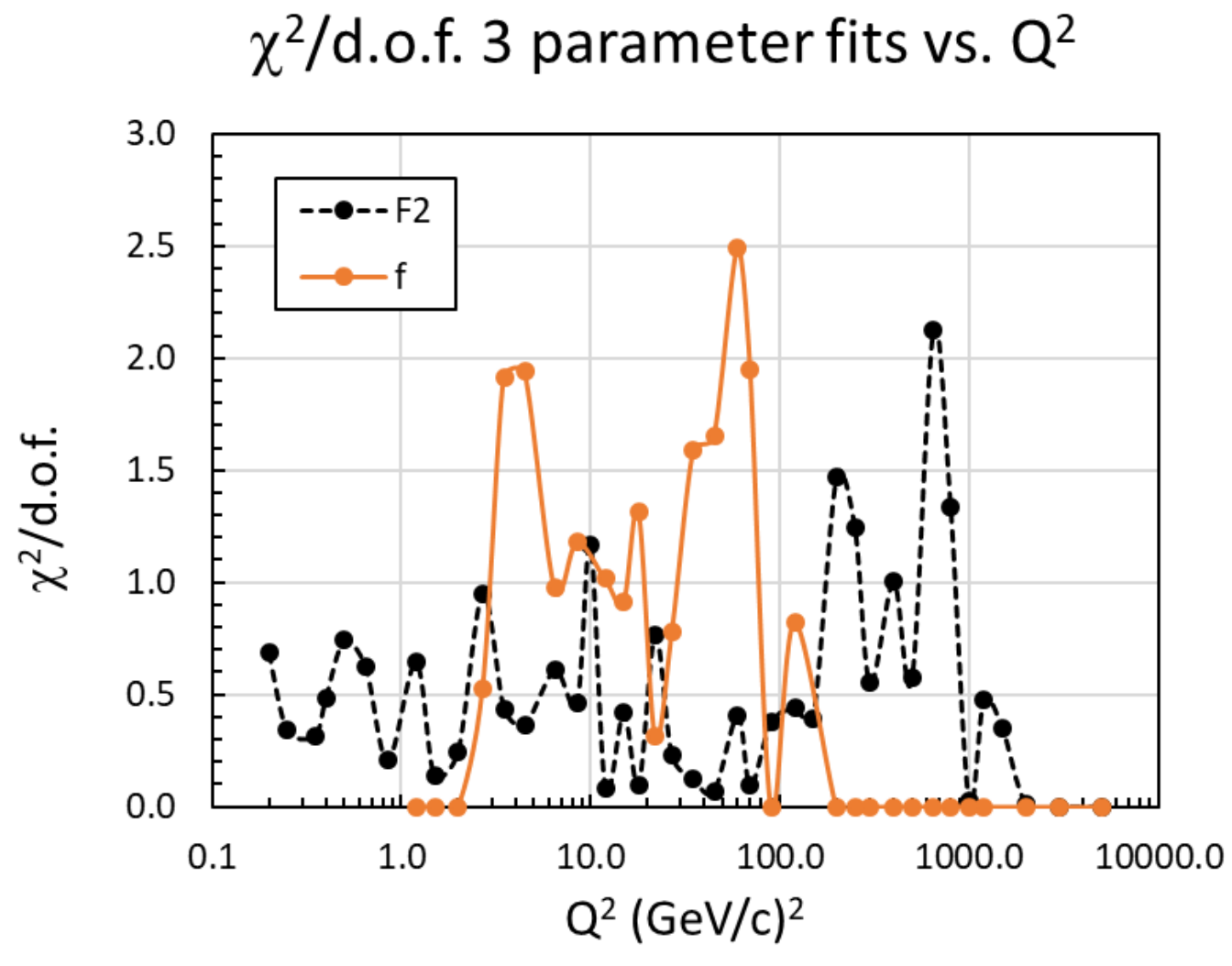

FIG. 3: The $\chi^2$/d.o.f. for 3 parameter fits of the reduced cross section data in Table 1 [4]. The points where the $\chi^2$/d.o.f. is zero correspond to 0-constraint fits where there were only 3 points of data.

The systematic corrections of the method were determined by computing the ratio of the true values (T) of $F_2$ and of $2xF_1$ at $x_{min}$ of the assumed $F_2$, $F_L$ model versus their respective fitted values (F). To model the T/F corrections the CTEQ structure function parameterizations were used under various assumptions of the magnitude of the gluon distribution evaluated through Eq. 4. As in the data analysis, minimum $\chi^2$ fits were performed on the simulated reduced cross section for each fixed valued of $Q^2$ as a function of x with the same average experimental errors and the same x-y values of the data.

## A. Estimation of $F_2(Q^2/s, Q^2)$

The determination of $F_2$ at $x_{min}$ for each Q value is nearly independent of the assumption of $F_L$ because the $y < y_{cut}$ (where $y_{cut} = 0.3$ to 0.5) made on the data for the fits suppresses the $F_L$ term by the $y^2/Y_+$ factor given in Eq. 1. For example, for $y_{cut} = 0.3$ the suppression of $F_L$ is < 1/16. Further, there are many parameterizations of $F_2$ that can be compared with our determination of $F_2$ at $x_{min}$. Nevertheless, $F_L$ makes a small contribution to the determination of $F_2$ so several assumptions for the $F_L$ are considered. These are labeled 'full' and 'half' glue corresponding to two assumptions for the gluon distribution based on the CTEQ parameterization and the "(4.7,0)" model that sets the gluon contribution in Eq. 4 to zero, but amplifies the $F_2$ contribution by a factor 4.7. Notice that the correction for $F_2$ is close to 1 for low Q but becomes of order 1.2 (20%) for high Q. The T/F correction is 'jagged' because it depends of the particular x values of the data. After all these considerations, we find $F_2$ is quite insensitive to various models of $F_L$ as is evident in Fig. 4.

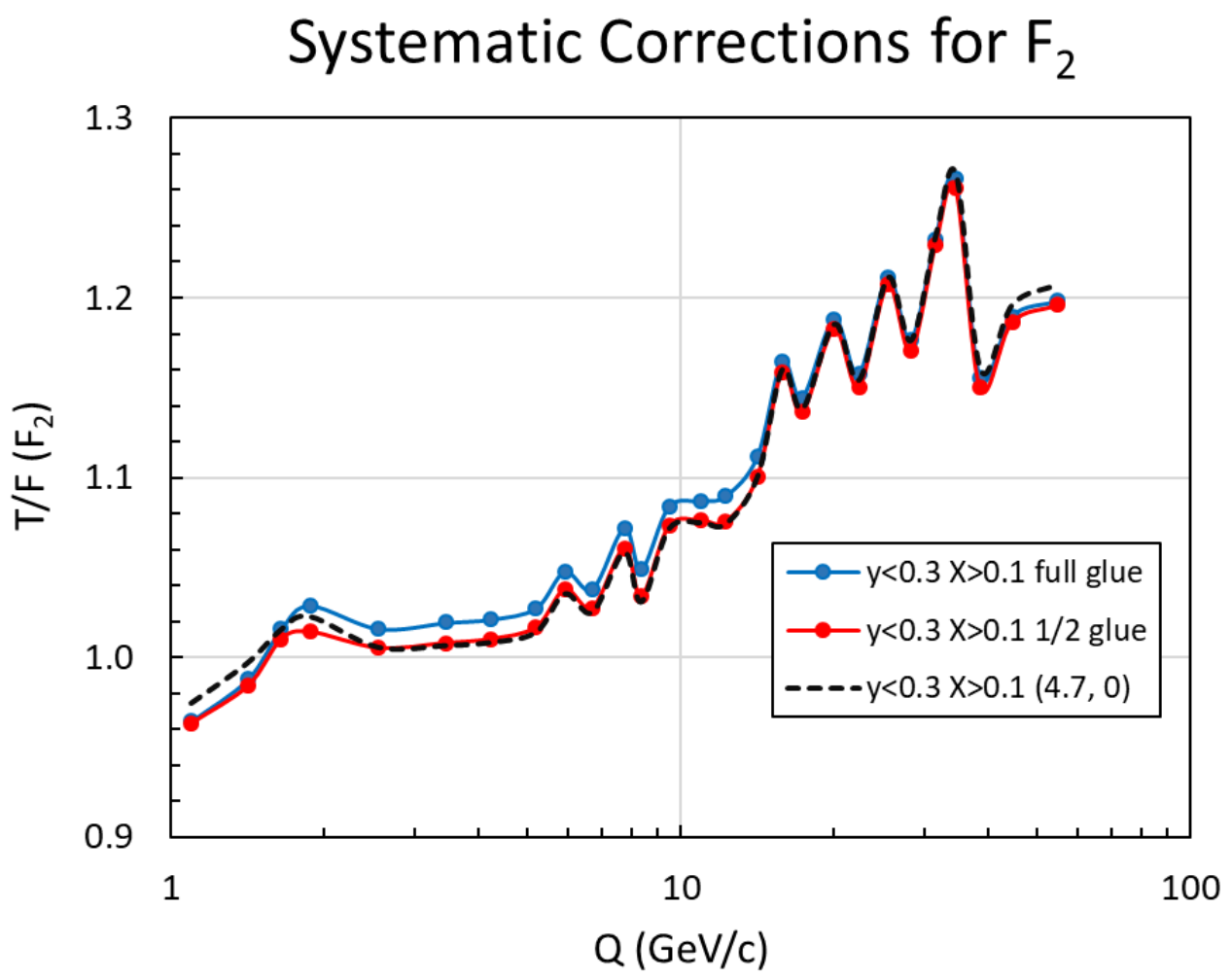

FIG. 4: Shown is the True/Fit (T/F) corrections for the determination of $F_2$ at $x_{min}$ based on a calculation of $F_2$ from the CTEQ parameterizations.

In Fig. 5 we plot $F_2$ at $x_{min} = Q^2/s$ versus $Q^2$ for Table 1 HERA data [4] compared to several parameterizations of $F_2$. We see that the data follows the expectation. The points on the graph are straight averages of four analyses of the reduced cross section. We employed two y-cuts, $y < 0.3$ and $y < 0.5$ each with two fitting functions - a three-parameter and a four-parameter polynomial fit. Each fit was corrected by its own truth/fit (T/F) value. The salient feature of the $F_2$ vs. $Q^2$ behavior is the rise for $Q^2 < 20$ $(GeV/c)^2$, the plateau from $Q^2 \sim 20$ to 100 $(GeV/c)^2$, then the subsequent fall for larger values of $Q^2$. The QCD evolution of the structure function for increasing Q as well as the fall off with increasing x determines the shape. At low $Q^2$ the former dominates, while the latter effect controls the larger $Q^2$ behavior. The data are compared to Abt et al. [10], CTEQ [25] and ALLM [26]. The latter two parameterizations are within ± 10% of the measured $F_2$ values, whereas the Abt. et al. parameterization developed from the HERA data themselves more closely follows the $F_2$ at xmin up to $Q^2 = 300$ (GeV/c)$^2$.

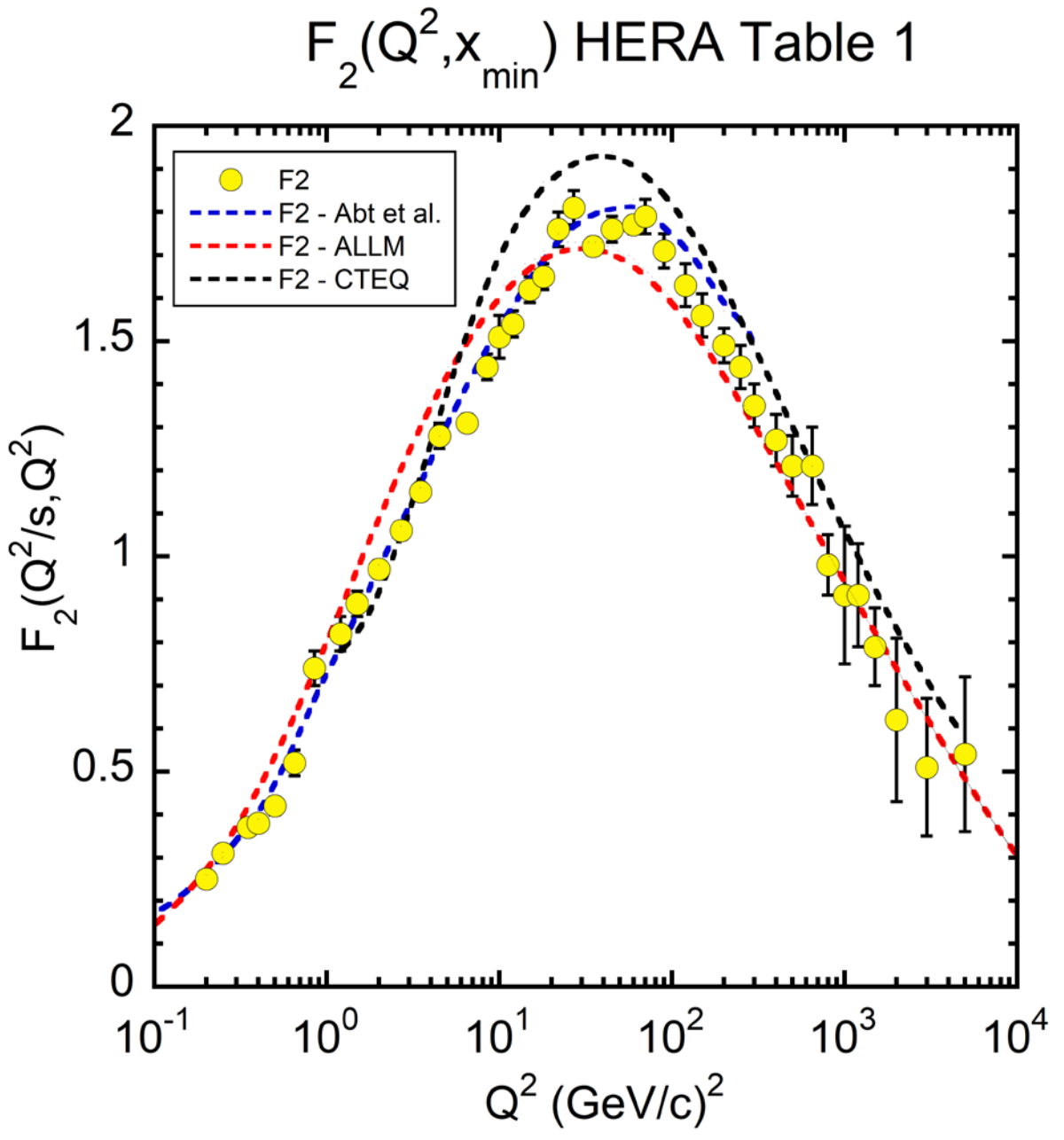

FIG. 5: $F_2$ at $x_{min}$ function is plotted (yellow circles) as a function of $Q^2$ for the √s = 318 GeV HERA Table 1 [4] data. Shown in comparison are the parameterizations of Abt et al. [10] blue dotted line), ALLM [26] (red dotted line), and CTEQ [25] (black dotted line).

$F_2$ at low $Q^2$ and $x_{min}$ should be sensitive to quark-antiquark thresholds. For DIS kinematics the mass squared of the virtual photon is negative, but when its value is small there is a probability that the virtual photon can fluctuate into a virtual quark-antiquark pair. Further, at y = 1 the exchanged photon is transversely polarized allowing a coupling to vector mesons.

Thus, we examine the behavior of $F_2$ quite closely in the low Q region. Interestingly, we find that $F_2(x_{min})$ determination at low $Q^2 \sim 1$ $(GeV/c)^2$ shows an anomaly that appears to be consistent with ϕ meson photoproduction threshold[1]. This behavior of $F_2$ from HERA data Table 1 [4] is shown in Fig. 6. In order to study this anomaly, we describe the $Q^2$ behavior above Q ~ 1 GeV/c – then compare this functional dependence to the $F_2$ values below Q ~ 1 GeV/c. We find that the $Q^2$ behavior of $F_2$ in the region up to $1 < Q^2 \leq 27$ $(GeV/c)^2$ follows a logarithmic dependence of the form $F_2(Q^2/s,Q^2) = (0.318 \pm 0.006) \times \ln(Q^2) + 0.76 \pm 0.01$ ($\chi^2$/d.o.f. = 0.62 for 12 d.o.f., p = 0.83). As another method of isolating the anomaly we fit $F_2$ at $x_{min}$ for the entire data set of Table 1 above $Q^2 \geq 1$ $(GeV/c)^2$ with a cubic of the form $\ln(F_2) = a \ln(Q^2)^3 + b \ln(Q^2)^2 + c \ln(Q^2) + d$ ($\chi^2$/d.o.f. = 0.99 for 31 d.o.f., p = 0.48). Finally, we find that if we fit the entire $F_2$ with no $Q^2$ cut to the cubic form the $\chi^2$/d.o.f. = 1.69 for 37 d.o.f. with p = 0.0054 indicating a poor fit. Hence the anomaly at Q ~ 1 GeV/c is statistically significant.

We show the threshold for these two fitting procedures for Q > 1 GeV/c in the two lower panels by the jump in the black dotted lines below and above Q ~ 1 GeV/c in Fig. 6. It is evident that the Abt et al. analysis tracks the discontinuity. The black dotted lines in the lower panel figures indicate the transition at Q ~ 1 GeV of roughly 20%. As a measure of the transition, we take the reciprocal of the $F_2$/Fit ratio for the largest Q point below the threshold as a measure of the threshold anomaly. The threshold around Q ~ 1 GeV/c is apparent. For the linear fit we find that the reciprocal equals 1.20 ± 0.07 and for the cubic case 1.19 ± 0.07.

[1] This low $Q^2$ region ~ 1 $(GeV/c)^2$ was highlighted in the paper by Abt et al. [10] who indicate that the $Q^2$ evolution behavior of the reduced cross section in this region shows "no abrupt change". This is true for $F_2$ but it is the first derivative of $F_2$ w.r.t. $\ln(Q^2)$ when plotted at $x_{min}$ that shows the anomaly.

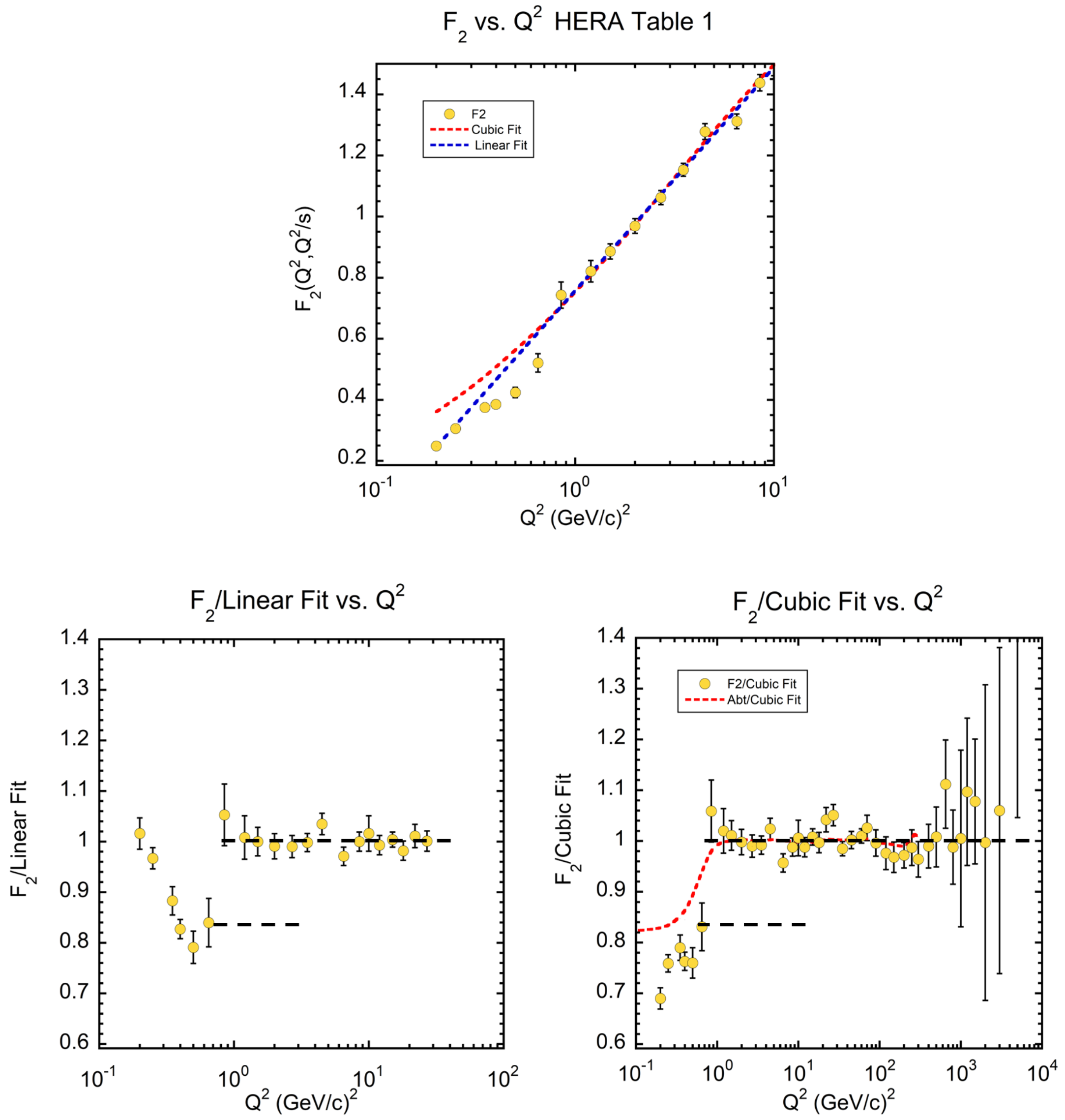

FIG. 6: The top panel shows $F_2$ at $x_{min}$ as a function of $Q^2$ down to very low $Q^2$ corresponding to x ~ few x $10^{-6}$. The red line corresponds to a linear fit to $F_2$ vs. $\ln(Q^2)$ and the blue dotted line indicates a cubic fit to $\ln(F_2)$ vs. $\ln(Q^2)$. In the lower left we show the ratio of the data to the linear fit emphasizing the discontinuity and in the lower right figure, we show the ratio of the $F_2$ data to the cubic fit. The red dotted line in the lower right panel is the result analyzing the Abt et al. parameterization in the same manner to a cubic polynomial in $\ln(Q^2)$.

From the form of the $F_2$ structure function, which is determined by the sum of the quark distributions - each coupled to the photon by their respective charge-squared values, we would naively expect the magnitude of the (u,d) → (u,d,s) transition to be:

$$\frac{F_2(Q^2 > 1\ (GeV/c)^2)}{F_2(Q^2 < 1\ (GeV/c)^2)} \approx \frac{\sum_{i=1}^{6} e_i^2}{\sum_{i=1}^{4} e_i^2} = 1.2, \qquad (6)$$

in agreement with the two experimental measures described above. We have assumed that the strange sea $xs = x\overline{s} = \left(x\overline{u} + x\overline{d}\right)/2$. Thus, the anomaly in $F_2$ evaluated at $x_{min}$ around Q ~ 1 GeV/c is consistent with the $s\overline{s}$ threshold in ϕ-meson photoproduction. This finding suggests that the dipole model of $F_2$ structure function is favored as suggested by Abt et al. [10] and others.

## B. Estimation of $2xF_1(Q^2/s,Q^2)$

For estimating the $2xF_1$ value at $x_{min}$, leading to the determination of $F_L$ itself, the dependence on the model of $F_L$ for the fit correction is much stronger. Various assumptions of $F_L$ were explored, each leading to its own systematic correction term, T/F. We show typical corrections (T/F) for three different assumptions of $F_L$ in Fig. 7. The T/F ratio is 'jagged' because it depends on the particular ensemble of x points of the data. The T/F correction is less than 1 indicating that $2xF_1$ uncorrected has to be corrected downward.

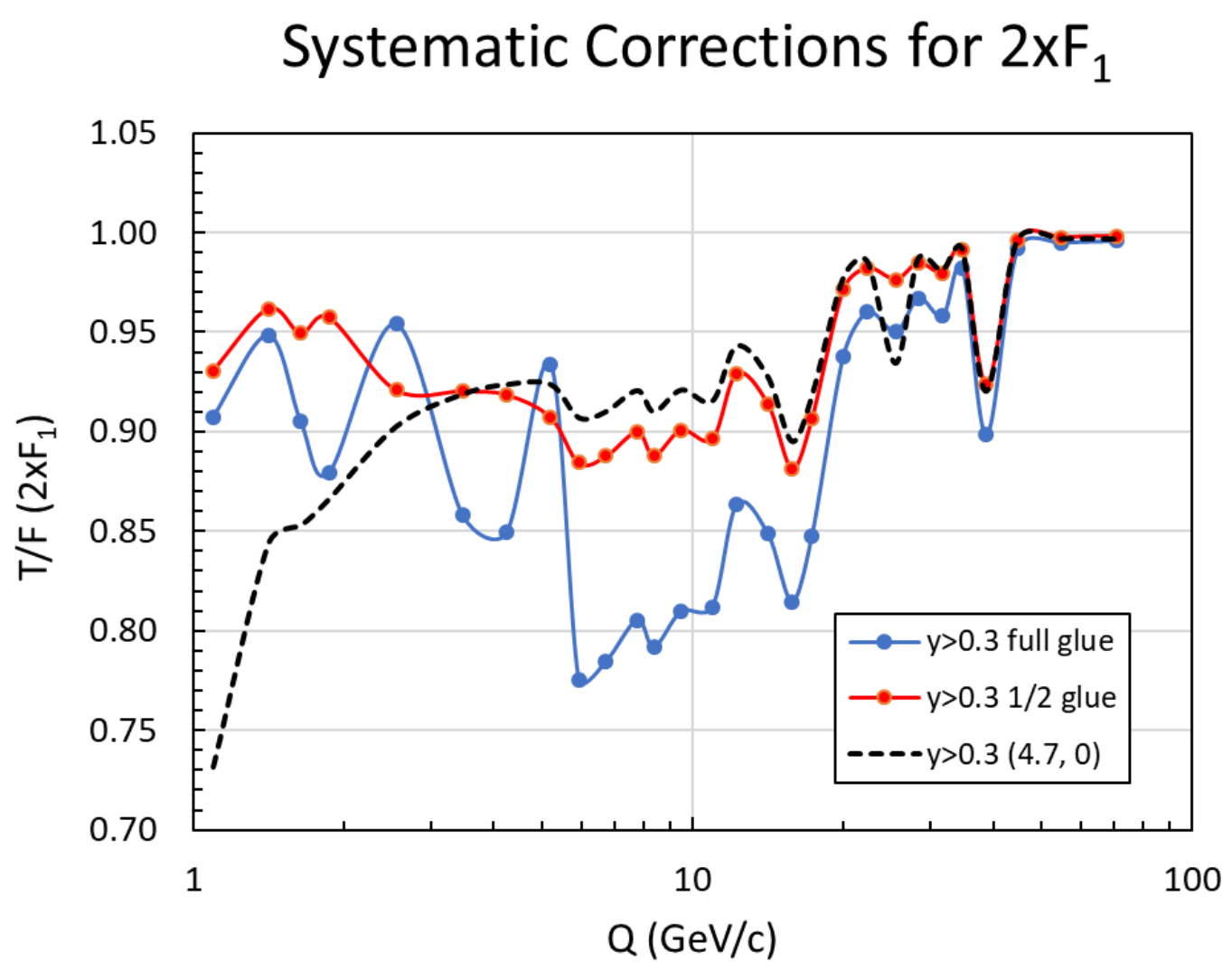

FIG. 7: Shown are the T/F ratios for the determination of $2xF_1$ as a function of $Q^2$ for various assumptions of $F_L$ shown by full glue/CTEQ blue-solid line, ½ glue/CTEQ red-solid line and no glue (4.7,0), but with the $F_2$ component in Eq. 4 enhanced by a factor of 4.7 as the black-dotted line. The black-dotted line is the preferred correction.

In order to converge on the best value of this crucial T/F correction for $2xF_1$ we consider the measured ratio $2xF_1/F_2$ of the data after correction compared to the model calculation of $2xF_1/F_2$. When the two plots converge, we have estimated the appropriate correction. Since the shape of $F_L$ is not a priori known, the $F_L$ function in the correction model was roughly tuned by determining the ratio $2xF_1/F_2$ for different relative magnitudes of the $F_2$ component versus the gluon xG component as a function of $Q^2$ in Eq. 4.

Several calculations of the ratio $2xF_1/F_2$ for different model assumptions are shown for the HERA Table 1 data [4] in Figs. 8. Fig. 8a we plot the comparison of data with models involving of the full gluon contribution given by the CTEQ [25] default values and one assuming the gluon component is ½ of its default value. In both these cases the gluon component dominates the $F_2$ component in Eq. 4. In Fig. 8b we have turned off the gluon component completely and enhanced the $F_2$ component to match the $2xF_1/F_2$ ratio of the data. In studying the two Figs. 8, it is obvious that the data strongly favor the no-gluon – enhance $F_2$ case over the default gluon or ½ gluon case. This conclusion is checked by a minimum $\chi^2$ fit to the $2xF_1/F_2$ ratio to determine the best enhancement factor of the $F_2$ component. The fit yields $F_2 \rightarrow (4.7 \pm 0.5)\ F_2$, with $\chi^2$/d.o.f. = 1.62 for 27 d.o.f. (p = 0.02). In general terms, the gluon component of $F_L$ rapidly goes to a small value as Q decreases in disagreement with data, whereas the $F_2$ component goes to a small value at a much slower rate and matches the measured $2xF_1/F_2$ ratio much better. Both the CTEQ $F_2$ (black dotted line) model and the Abt et al. model (red-dotted line) agree with this conclusion as seen in Fig. 8b.

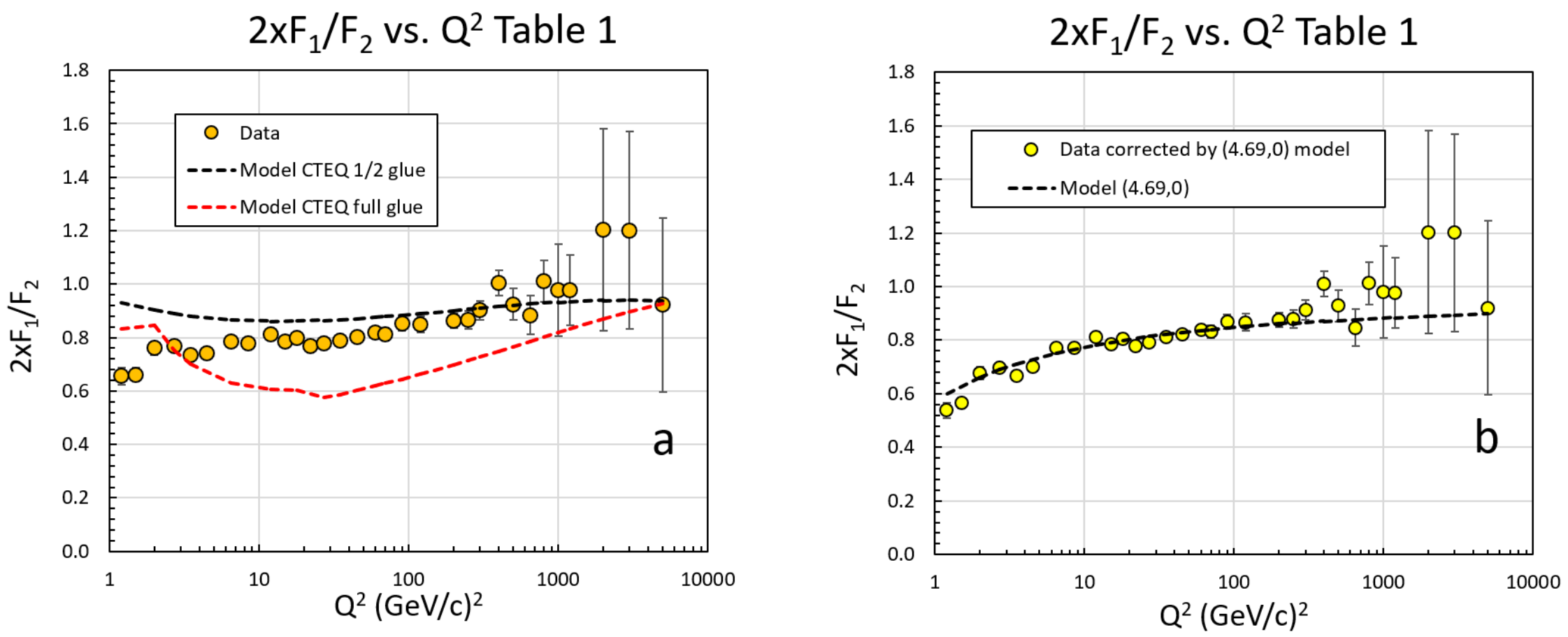

FIG. 8: The $2xF_1/F_2$ ratio as a function of $Q^2$ for Table 1 HERA combined data is shown for two different models of $F_L$. In Fig. 8a the black dotted line is the estimated $2xF_1/F_2$ ratio from the CTEQ parameterization of one half the CTEQ gluon default value and the red-dotted line represents the $2xF_1/F_2$ value of the full gluon value. The data points in the figure have been corrected by the ½ gluon case. In Fig. 8b shows the $2xF_1/F_2$ ratio after tuning the relative size of the $F_2$ component to the xG component of the $F_L$ term.

## C. Estimation of systematic errors

Many comparisons of the estimated values of $F_2$ and $2xF_1$ were made to check that the systematic effects associated with different cuts and fits were properly corrected. Among many tests of the systematic errors, different regions of y were used, such as $y < 0.3$ ranging to $y < 0.5$, for the $F_2$ determination and $y > 0.3$ to $y > 0.5$ for the $2xF_1$-estimate. In addition, a cut ($\ln(\ln(1/x)) > 0.1$ (corresponding to $x < 0.33$) was made to avoid fitting the high x-region of the $F_2$ structure functions where the valence quarks contribute. In cases where the extent of the data permitted, fits were performed with a cubic vs. quadratic polynomials in the double log x variable. It was found that this ensemble of analyses was consistent within ~ ± 10% - therefore the results of different fits for a given s were averaged with equal weights as were their respective statistical errors for the best determination of $F_2$ and $2xF_1$ at $x_{min}$.

Since Table 1 of the HERA data set [4] has the most extensive data, we performed most of our systematic tests there. Table I below lists the chief systematic errors considered for HERA data sets of Tables 1, 3 and 4.

TABLE I. Shown are the systematic errors for various changes in the analysis and corrections for HERA Tables 1, 3 and 4 where both $F_2$ and $2xF_1$ are determined. The CTEQ parameterizations were used to estimate $F_2$ and $2xF_1$ with the same acceptance cuts as the data.

| Function | Systematic Change | Result |
|---|---|---|
| $F_2$ | $y < 0.3$ vs. $y < 0.5$ $F_L$ evaluated with default CTEQ [25] gluon distribution | $-5\% \leq \Delta F_2/F_2 \leq 12\%$ |
| $F_2$ | $y < 0.5$ $F_L$ evaluated with default CTEQ vs. ½ default gluon distribution | $\Delta F_2/F_2 < 4\%$ |
| $F_2$ | For $y < 0.3$ difference of three vs. four parameter fits | $\lvert\Delta F_2/F_2\rvert \sim 7\%$ |
| $2xF_1$ | $y > 0.3$ full glue vs. ½ glue | $-12\% \leq \Delta F_1/F_1 \leq 4\%$ |
| $2xF_1$ | $y > 0.3$ vs. $y > 0.5$ for three-parameter fit | $\lvert\Delta(F_1)/F_1\rvert \sim 5\%$ |

And the estimated systematic errors for the $2xF_1$ determination of data of Table 2 are estimated below.

TABLE II. Shown are the systematic errors for various changes in the analysis and corrections for Table 2 where only $2xF_1$ is determined from data. The $F_2$ systematic is estimated from differences between the CTEQ and Abt et al. $F_2$ parameterizations.

| Function | Systematic Change | Result |
|---|---|---|
| $F_2$ | CTEQ[25] vs. Abt et al. [10] $F_2$ Model | $-10\% \leq \Delta F_2/F_2 \leq 10\%$ |
| $2xF_1$ | y>0.3 two point vs. three point fits | $-7\% \leq \Delta F_1/F_1 \leq 4\%$ |
| $2xF_1$ | y>0.3 full vs. ½ glue correction two point fits | $-6\% \leq \Delta F_1/F_1$ |

From examination of the two tables we would roughly estimate that the systematic error on $F_2$ is ~ ± 10% and the corresponding error for $2xF_1$ is ~ ± 10%. Thus, the systematic error in $F_L$, derived from the subtraction of these two structure functions with errors largely uncorrelated is ≤ ± 15%.

## III. THE $F_2$ AND $2xF_1$ DISTRIBUTIONS

Before we delve into the properties of $F_L$, it is interesting to study the behavior of the $F_2$ and $2xF_1$ distributions at $x = Q^2/s$ themselves. In Fig. 9 we show $F_2$ and $2xF_1$ as a function of $Q^2$ for each of the four center-of-mass energies of the HERA combined reduced cross section data [4]. Remember that it is the difference between these two functions that is the longitudinal structure function, $F_L$. Its nonzero value, given by the difference between the black circles and red points triangles in the figure, is obvious from the figures. It is apparent that $F_2 > 2xF_1$ for low $Q^2$ but at $Q^2 \sim 10^2$ $(GeV/c)^2$ the two become equal. Thus, the shape of $F_L$ as a function of $Q^2$ can be easily visualized. For Table 2 data shown in the upper right panel, the red triangles are for two point fits, the yellow triangles are fits with greater than two points and the black dotted line represents $F_2$ from a parameterization.

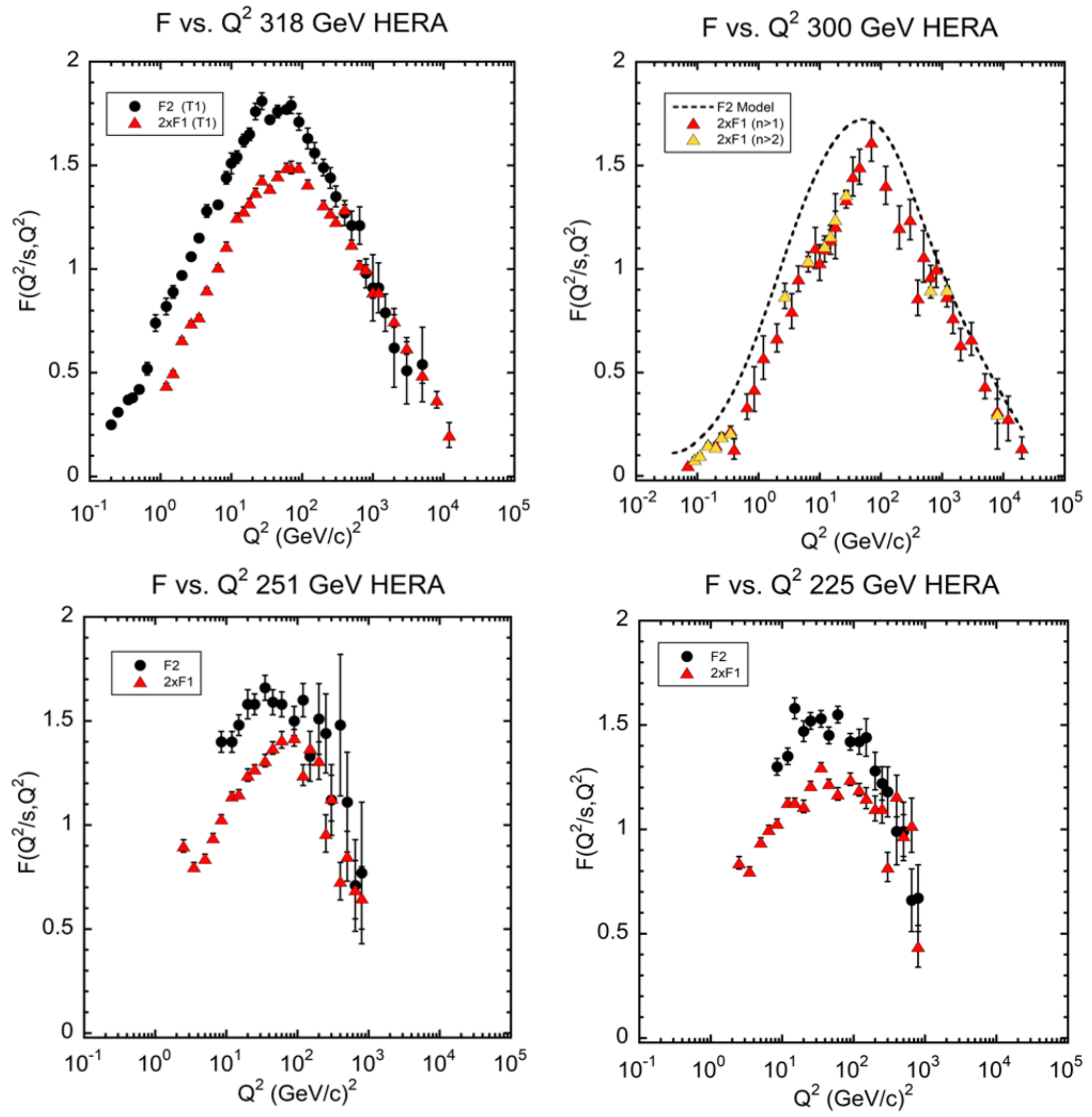

FIG. 9 Shown are the $F_2$ (black circles) and $2xF_1 = F_2 - F_L$ (red triangles) structure functions determined at $x_{min}$ as a function of $Q^2$ for tables ($\sqrt{s}$ = 318, 300, 252, 225 GeV) of the HERA combined data [4]. Only the statistical errors are shown.

The salient feature of both $F_2$ and $2xF_1$ distributions at $x_{min}$ is that they peak at certain value of Q that is s-dependent. The peak is where $dF_2/d\ln(Q^2)$ changes sign. In order to determine the best values of the peaks in both the $F_2$ and $2xF_1$ distributions, we deployed a minimum $\chi^2$ fit of a gaussian in the natural logarithm of $Q^2$:

$$F(Q^2, x_{\min}) = F_0 \exp\left( \frac{(d - P_1)^2}{P_2^2} \right), \qquad (7)$$

where F is either $F_2$ or $2xF_1$, $F_0$ is the peak value of the F distribution, $d = \ln(Q^2)$, $P_1$, which equals $\ln(Q_0^2)$, is the $\ln(Q^2)$ value at the peak and $P_2$ is the logarithmic width of the distribution. We avoided the very low $Q^2$ below the $\phi$ threshold by requiring that Q > 1 GeV/c.

The results of the fits are tabulated in Table III for $F_2$ at $x_{min}$. Notice that the peak and its $Q^2$ location, given by exponentiation of $P_1$, increases with increasing s. The logarithmic width, given by $P_2$, is roughly constant - independent of $\sqrt{s}$. The $\chi^2$/d.o.f. of the fits are generally good meaning that the function form in Eq. 7 is an adequate description of the data. Only the statistical errors were included.

TABLE III. The fit values for the $F_2$ at $x_{min}$ distributions are tabulated. There are no data for Table 2 since all measurements are at high y. For the $F_L$ analysis the data of Tables 3 and 4 were combined statistically.

| HERA Data Set | $\sqrt{s}$ (GeV) | $Q_0^2$ $(GeV/c)^2$ | $(F_2)_0$ | $P_2$ | $\chi^2$ | d.o.f. |
|---|---|---|---|---|---|---|
| Table 1 | 318 | 44.8 ± 1.3 | 1.75±0.01 | 4.0 ± 0.1 | 39.7 | 31 |
| Table 3 | 251 | 41.8 ± 4.1 | 1.61±0.03 | 3.8 ± 0.3 | 10.4 | 15 |
| Table 4 | 225 | 37.6 ± 2.6 | 1.53±0.02 | 3.7 ± 0.2 | 19.0 | 15 |
| Table 3&4 | 238 | 38.0 ± 2.2 | 1.56±0.01 | 3.8 ± 0.2 | 18.4 | 15 |

The $Q_0^2$ values of Table 3 can be converted to $x_{min}$ values of where $dF_2/d\ln(Q^2)$ changes sign yielding for $\sqrt{s}$ = 318, 251, 238, 225 GeV, $x_{min} = (4.4 \pm 0.1) \times 10^{-4}$, $(6.6 \pm 0.7) \times 10^{-4}$, $(6.7 \pm 0.4) \times 10^{-4}$, $(7.4 \pm 0.5) \times 10^{-4}$, respectively. These values are smaller than were $dF_2/d\ln(Q^2)$ changes sign in the Abt. et al. paper [10] ($x \sim 5 \times 10^{-3}$). Fitting the Abt et al. [10] parameterization at $x_{min}$ with the same function yields approximately the same values.

In the following table (Table IV) we show the fit parameters for the $2xF_1$ functions at $x_{min}$. Notice that the $\chi^2$ of these fits is generally poor, where we find that lowest Q point of each distribution contributes approximately ½ of the $\chi^2$ of each fit, but overall the distributions are "noisy". Adding in quadrature a systematic error of 10% reduced the $\chi^2$ for Table 1 to 6.4 for 30 d.o.f. and for Table 3 the $\chi^2$ = 14.5 for 19 d.o.f. with fit values within errors.

TABLE IV. The fit values for the $2xF_1$ at $x_{min}$ distributions are tabulated.

| HERA Data Set | √s (GeV) | $Q_0^2$ $(GeV/c)^2$ | $(2xF_1)_0$ | $P_2$ | $\chi^2$ | d.o.f. |
|---|---|---|---|---|---|---|
| Table 1 | 318 | 69.2±1.0 | 1.48±0.01 | 3.83±0.02 | 121.7 | 30 |
| Table 2 | 300 | 61.6±3.2 | 1.38±0.02 | 4.1±0.1 | 39.4 | 37 |
| Table 3 | 251 | 56.9±3.7 | 1.33±0.01 | 3.9±0.1 | 99.9 | 19 |
| Table 4 | 225 | 46.5±2.6 | 1.24±0.01 | 4.2±0.1 | 66.7 | 19 |
| Table 3&4 | 238 | 49.9±2.1 | 1.27±0.01 | 4.0±0.1 | 84.6 | 19 |

# III. $F_L$ DATA

Our analysis of $F_L$ involves the determination of two statistically independent functions: $2xF_1 = F_2 - F_L$ determined from the high-y shape of the reduced cross section; and $F_2$ extracted from the lower-y x-dependence of the reduced cross section. The summary of the data used is shown in Table V. The corresponding $x_{min}$ ranges of the data may be simply evaluated by $x_{min}=Q^2/s$.

TABLE V. The HERA combined H1 and ZEUS reduced cross section data sets and their respective kinematic regions for this analysis are shown.

| HERA Table Number | √s (GeV) | Electron Beam | Kinematic Range $F_2$ $(GeV/c)^2$ | Kinematic Range $2xF_1$ $(GeV/c)^2$ |
|---|---|---|---|---|
| 1 | 318 | $e^+$ | $0.2 < Q^2 < 5000$ | $1.2 < Q^2 < 12000$ |
| 5 | 318 | $e^-$ | $150 < Q^2 < 3000$ | $90 < Q^2 < 12000$ |
| 2 | 300 | $e^+$ | | $0.65 < Q^2 < 20000$ |
| 3 & 4 | 238 | $e^+$ | $8.5 < Q^2 < 800$ | $2.5 < Q^2 < 800$ |

Table 1 has the most extensive data covering a wide region of $Q^2$ and hence $x_{min}$ ($2 \times 10^{-6} < x_{min} < 0.12$). Table 2 is largely at high y enabling a determination of $2xF_1$ to be made but not of $F_2$. The data of HERA Tables 3 and 4 were combined statistically. The data of HERA Table 2 had to be treated specially. Since all the data were at high y it was not possible to independently determine $F_2$. Therefore, we used the Abt et al. [10] parameterization for $Q^2 \leq 300$ $(GeV/c)^2$ and the CTEQ [25] parameterization normalized by a multiplicative factor of 0.97 to Abt et al. at $Q^2 = 300$ $(GeV/c)^2$ for larger values of $Q^2$. Another exception had to be made in analyzing the data of Table 2. There were a few values of $Q^2$ where there were three or more x-points existed which would have enabled at least a 0-C fit to a quadratic extrapolation to $x_{min}$. At the $Q^2$ values where there were only two x-points, we performed a linear fit to determine $2xF_1$ at $x_{min}$. In both fitting cases the T/F corrections were made with identical fitting procedures.

The resulting $F_L$ functions as a function of $Q^2$ are shown in Figs. 10. Here we distinguish between the determinations from the HERA tables where both $F_2$ and $2xF_1$ have been extracted from the data from the analysis of HERA Table 2 that had the exceptions noted above. Unlike other analyses of $F_L$, where a range of y < 1 values are included, our analysis is an evaluation of the $F_L$ function at y = 1 where the polarization of the virtual photon is all transverse. Hence, we are not strictly measuring $F_L$ (the "longitudinal" structure function) but rather the correction term to $F_2$ to make it equal to $2xF_1$.

It is in these figures where we encounter a surprise that was already broached in the discussion of the $2xF_1/F_2$ ratio above. We notice that the gluon-dependent models of $F_L$ as calculated by Eq. 4 are strongly disfavored – that is the black dotted curves of Fig.10 do not go through the low $Q^2$ data. This is quite different from other determinations of $F_L$ that indicated that the gluon component dominates. Thus, we would have expected that $F_L$ at $x_{min}$ will be small because the dominate gluon component goes to zero for transversely polarized virtual photons. The $F_L$ function at $x_{min}$ peaks at $Q^2 \sim 10$ (GeV/c)$^2$, but surprisingly remains finite even at low $Q^2$ where, if $F_L$ were dominated by gluons, it should be small. We have already seen this behavior in the shape of $2xF_1/F_2$ shown in Fig. 8b.

We compare the expectation for $F_L$ for several different models of the gluon and $F_2$ components in the region Q > 1.44 GeV/c where our fit to $\alpha_s(Q^2)$ from the PDG are valid. The Models 1, 2 curves shown in Fig. 10 are based on scaling the CTEQ parameterization of $F_2$ and xG evaluated at $x_{min}$. Model 1 scales the gluon contribution by $xG \rightarrow 0.35\ xG$ in Eq. 4 to follow the data above $Q^2 \sim 30$ (GeV/c)$^2$ but fails to describe the data below $Q^2 \sim 10$ (GeV/c)$^2$. Model 2 uses the CTEQ parameterization of $F_2$ by enhancing it by $F_2 \rightarrow (4.7 \pm 0.05)\ F_2$, but sets the gluon distribution completely to zero. Model 3 does the same enhancement of $F_2$ using the Abt et al. parameterization [10] again with the gluon contribution turned off. This behavior is consistent with expectations of $F_L$ at $x_{min}$. The gluon contribution to $F_L$ is a measure of the longitudinal virtual photon component, but at $x_{min}$ where the virtual photon is transverse, this component is strongly disfavored.

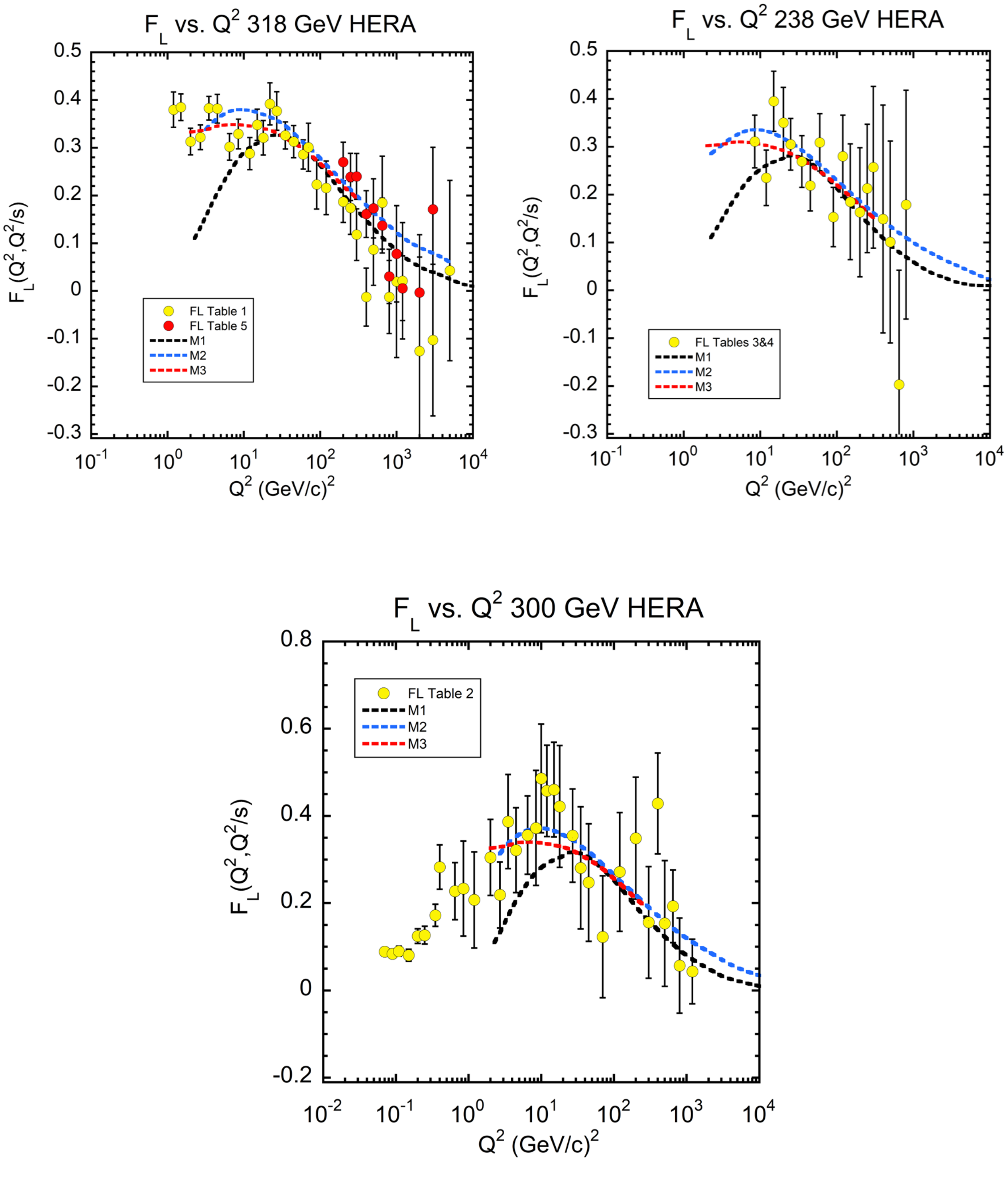

FIG. 10: The extracted longitudinal structure function is shown as a function of $Q^2$ for three values of $\sqrt{s}$. Only statistical errors are shown. The data are compared to three models of $F_L$ evaluated by Eq 4 for Q > 1.44 GeV/c.

As a measure of the conclusion that the gluon component of $F_L$ at $x_{min}$ is strongly suppressed, we compute the $\chi^2$/d.o.f. of each of the three models for the data of Table 1. For Model 1, which is an evaluation of the CTEQ PDFs through Eq. 4 with the default gluon suppressed by a multiplicative factor of 0.35, the $\chi^2$/d.o.f. = 7.7 for 34 d.o.f. - hence strongly disfavored. For Model 2, that assumes the CTEQ $F_2$ component enhanced by a factor of 4.7 but no gluon contribution, the $\chi^2$/d.o.f. = 1.7 for 34 d.o.f. and Model 3 that uses the Abt. et al. $F_2$ parameterization enhanced by 4.7 for the $F_L$ calculation in Eq. 4 but with no gluon contribution, the $\chi^2$/d.o.f. = 1.3 for 17 d.o.f. Hence, the data strongly disfavors the CTEQ default gluon component of Model 1. These results and the results of comparing the three models with Tables 2 and 3&4 are shown in Table VI.

TABLE VI. The $\chi^2$ measure of the compatibility of different models to the HERA FL data for Q > 1.45 GeV to the high Q limit of the model. The HERA Data Sets 1 and 5 are at $\sqrt{s}$ = 318 GeV, Data Set 2 at 300 GeV and Data Set 3&4 are the statistically combined Sets 3 (251 GeV) and 4 (225 GeV) [4]. Model 1, which has a large gluon component, is strongly disfavored for Data Set 1+5. Models 2 and 3 yield better $\chi^2$ and are of roughly the same reasonably good quality for all data sets.

| Model | HERA Data Set | $\chi^2$/d.o.f. | d.o.f. |
|---|---|---|---|
| 1 | 1 + 5 | 7.7 | 34 |
| | 2 | 1.7 | 27 |
| | 3&4 | 1.3 | 12 |
| 2 | 1 + 5 | 1.7 | 34 |
| | 2 | 0.79 | 27 |
| | 3&4 | 0.95 | 12 |
| 3 | 1 + 5 | 1.3 | 17 |
| | 2 | 0.86 | 17 |
| | 3&4 | 1.1 | 8 |

In order to see the overall behavior of $F_L$ at $x_{min}$ we plot all the data on the same graph in Figs. 11a,b. The data are not good enough to measure the $\sqrt{s}$ - dependence of $F_L$ at $x_{min}$ although we notice that the points tend to coalesce for $x_{min} > 1 \times 10^{-3}$.

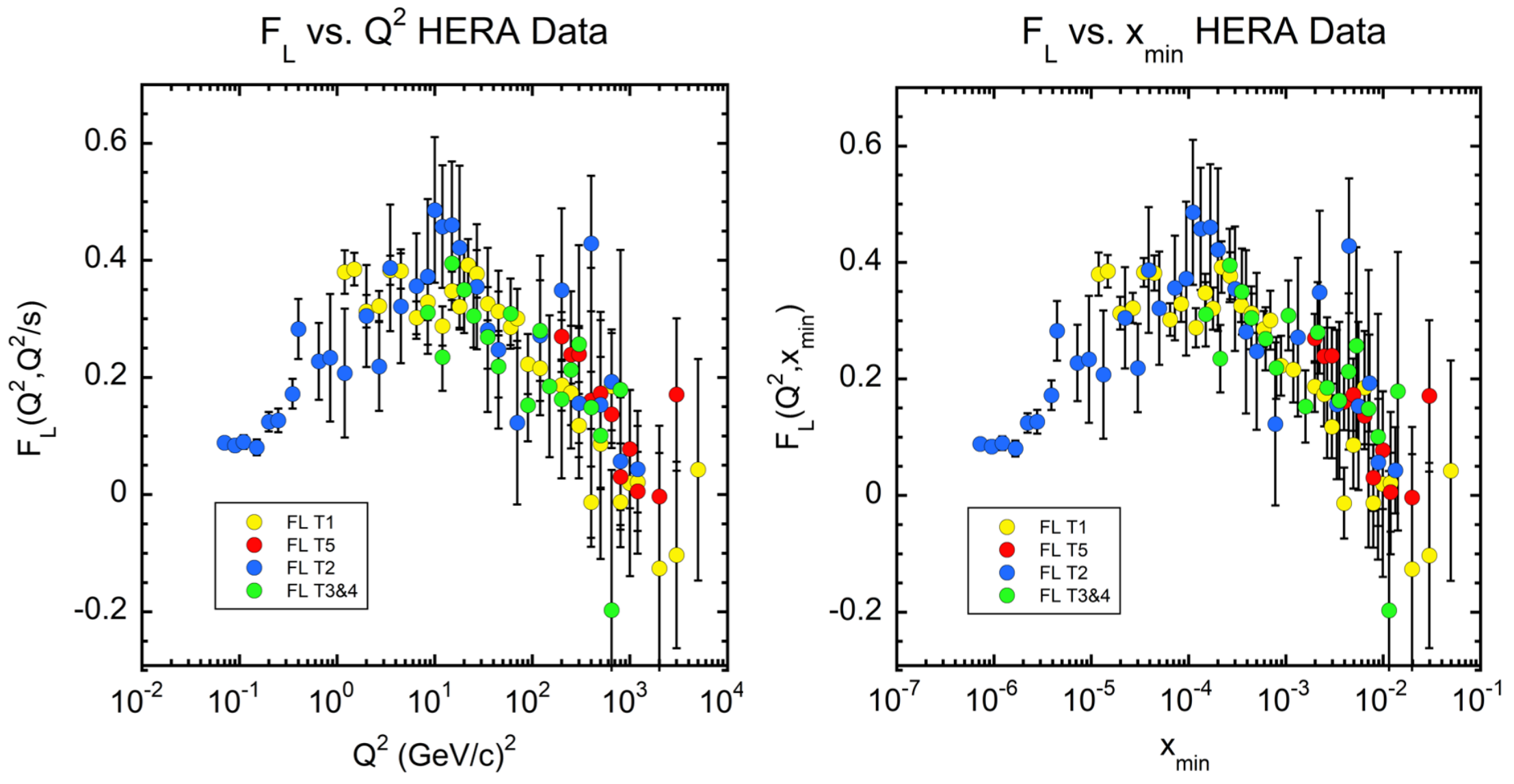

FIG. 11: Plotted are the $F_L$ values of our analysis at $x_{min}$ for all the NC DIS ep data of the combined H1-ZEUS HERA data sets [4] from $225 \leq \sqrt{s} \leq 318$ GeV. On the left, we plot $F_L$ as a function of $Q^2$ and on the right as a function of $x_{min} = Q^2/s$. Only the statistical errors are shown, but the systematic errors are roughly 15%.

As in the fits of $F_2$ and $2xF_1$ of Fig. 9, we fit the cumulative $F_L$ data with Eq. 7. The results are given in the Table VII below yielding the values of $Q_0^2$ and $x_{0min}$ where the distributions peak.

TABLE VII. The parameters of minimum $\chi^2$ fits to the form Eq. 7 for the combined $F_L$ data at $x_{min}$. The p-values for $Q^2$, $x_{min}$ fits are $9.3 \times 10^{-2}$, $7.2 \times 10^{-2}$, respectively.

| $Q^2$ fit | $Q_0^2$ (GeV/c)$^2$ | $F_0$ | $P_2$ | $\chi^2$ | d.o.f. |
|---|---|---|---|---|---|
| | 9.1±0.7 | 0.361±0.008 | 3.87±0.10 | 115.9 | 97 |
| $x_{min}$ fit | $x_{0min}$ | $F_0$ | $P_2$ | $\chi^2$ | d.o.f. |
| | (9.7±0.8) x $10^{-5}$ | 0.364±0.008 | 3.85±0.10 | 118.0 | 97 |

The ratio $R = F_L/(F_2 - F_L)$ at $x_{min} = Q^2/s$ is plotted in Fig. 12. Nominally, R is a measure of the longitudinal to transverse cross section ratio, but since the longitudinal polarization is zero at $y = 1$ this ratio becomes a characterization of the dip in the reduced cross section at $x_{min}$, $2xF_1$, compared to expectation $F_2$, so $R = F_2/2xF_1 - 1$. We have computed the value of R vs. $Q^2$ by Eq. 4 that is shown in Fig. 12 as the red-dotted curve, by enhancing the $F_2$ component by a factor 4.7 ± 0.5 and suppressing the gluon component completely. We note that R is independent of $\sqrt{s}$ when plotted vs. $Q^2$ since the QCD evolutions of both $F_L$ and $F_2$-$F_L$ are controlled by the evolution of $F_2$ itself. A computation of R with full QCD evolution using either the CTEQ [25] or the Abt et al. [10] parameterizations of $F_2$ confirms this conclusion. The red dotted curve is a reasonable representation of the data when tested with $\chi^2$/d.o.f. = 1.18 for 85 d.o.f. ($p = 0.12$). In our calculation of R, we have required that $Q^2 \geq 2$ $(GeV/c)^2$ to be in the range where the fit of $\alpha_s(Q^2)$ from the PDG [14] data is valid.

The $x_{min}$ region below $Q \sim 1.4$ GeV/c is interesting and imposes a challenge to simulation through the evaluation of $F_L$ by Eq. 4. From Fig. 12, it is clear that R continues to increase as $Q^2 \rightarrow 0$, becoming ~ 2 at $Q^2 \sim 0.1$ $(GeV/c)^2$. This is because $F_L$ remains larger that expectation when compared to $F_2$. In this low $Q^2$ region, we saw the anomaly in $F_2$, which we attributed to the $s\bar{s}$ threshold in $\phi$-meson photoproduction. Another difficulty in simulating R in this low Q region is the running of $\alpha_s(Q^2)$. By allowing $\alpha_s(Q^2)$ to continue to increase, the computed value of $F_L$ becomes so large that R diverges – a clearly unphysical situation, even though the $F_2$ Abt et al. parameterization [10] is still valid. The model of $F_L$ by Eq. 4 breaks down and some new "physics" is needed. Perhaps $\alpha_s(Q^2)$ "freezes" as $Q^2 \rightarrow 0$, as described by Deur, Brodsky and Roberts [27], or the dipole picture can step in to give a better picture of the relevant physics.

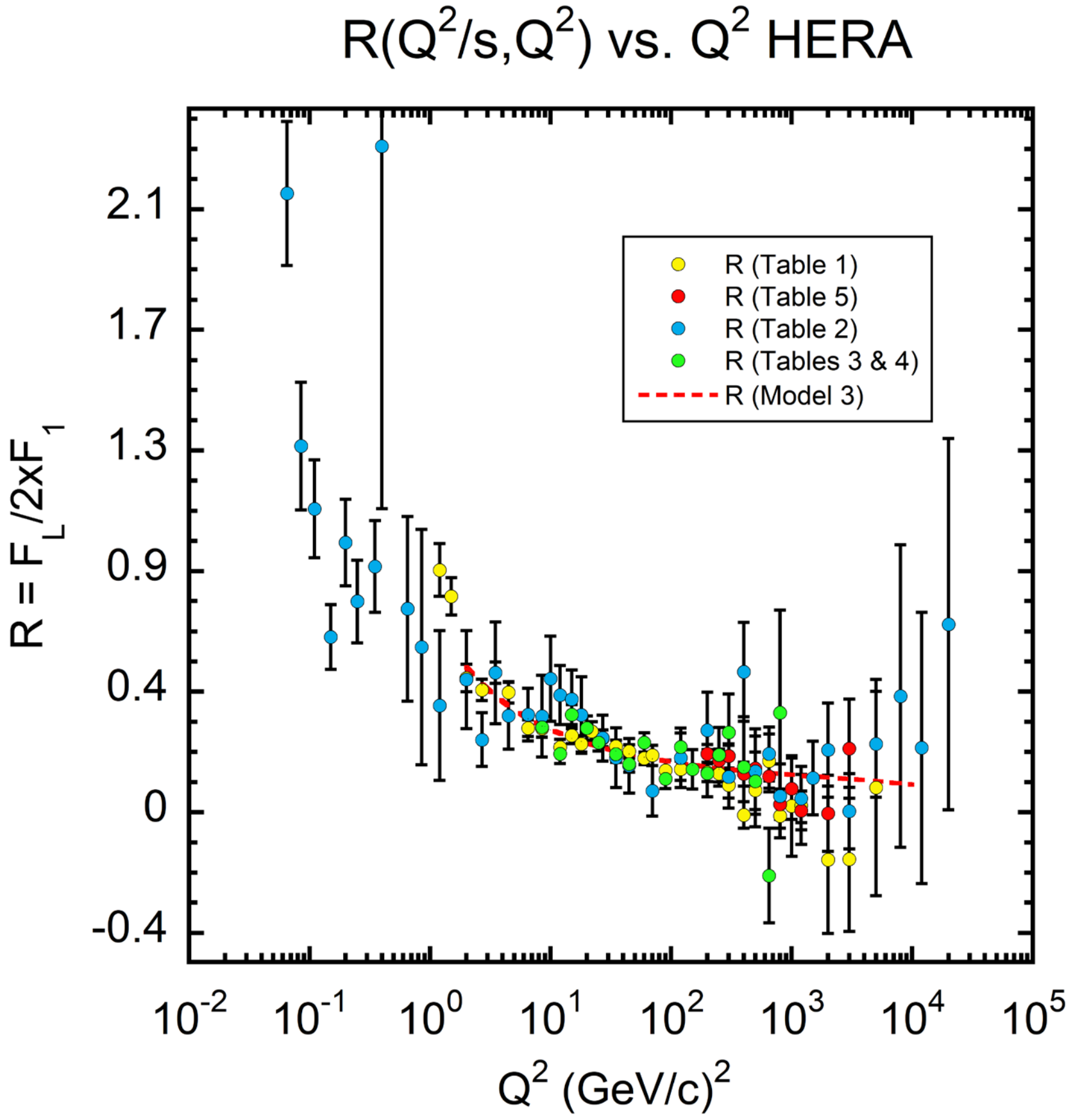

FIG. 12: We plot the ratio $R = F_L/2xF_1$ as a function of $Q^2$ for the HERA NC ep inclusive scattering data $225 \le \sqrt{s} \le 318$ GeV. The red-dotted line is a computation of R through $F_L$ evaluated by Eq. 4 for $Q^2 \ge 2$ (GeV/c)$^2$.

Both the H1 [15] and ZEUS [16] collaborations have determined $F_L$ by Rosenbluth analyses. Hence, their measurements are not directly comparable to our determination but they are closely related. In Fig. 13 we show the comparison. In the figure we represent our determination of $F_L$ at $x_{min}$ by a three parameter fit of the form given by Eq. 7 to the entire HERA data set with a $\chi^2$/d.o.f. = 1.2 for 97 d.o.f. (p = 7.2 x $10^{-2}$). Notice that the H1 and ZEUS $F_L$ determinations are smaller at small $Q^2$ than our $x_{min}$ analysis. This is consistent with our $x_{min}$ investigation that favors having $F_L$ controlled by $F_2$ with a zero gluon contribution, whereas the H1 and ZEUS determinations of $F_L$ include y < 1 data, where the longitudinal polarization is nonzero. This allows the gluon component of Eq. 4 to be operative. Note that the H1 and ZEUS determinations of $F_L$ go to zero as $Q^2 \to 0$ much faster than our determination of $F_L$ at $x_{min}$.

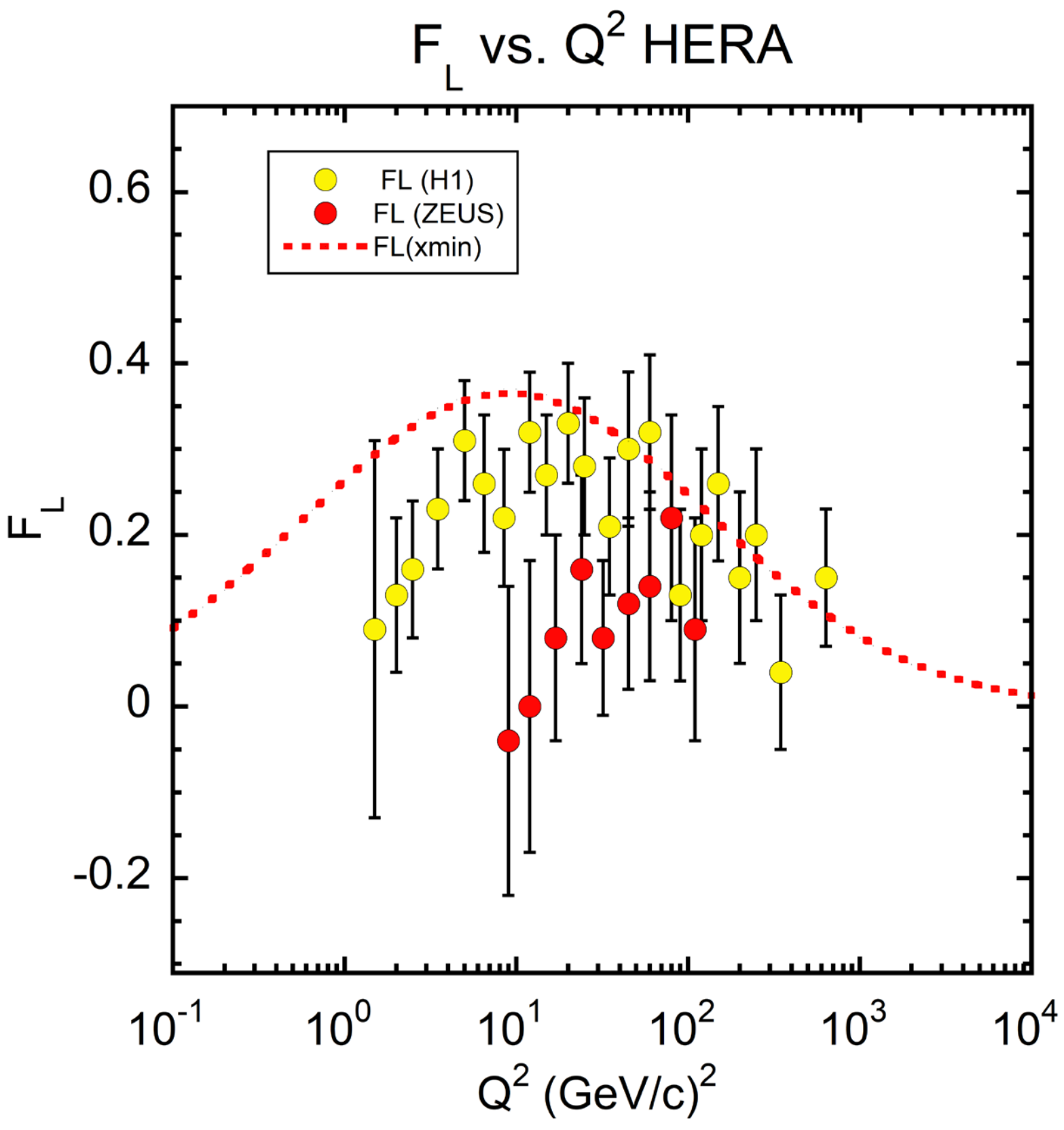

FIG. 13 The $F_L$ determinations of H1 and ZEUS collaborations are plotted as a function of $Q^2$. The red dotted line represents our determination of $F_L$ at $x_{min}$ in the range $7 \times 10^{-2} \leq Q^2 \leq 2 \times 10^4$ (GeV/c)$^2$.

Our determination of R (Fig. 12) and those of H1 and ZEUS (Fig. 13) become roughly equal for $Q^2 > 100$ (GeV/c)$^2$.

# IV. CONCLUSIONS

We have determined a large correction to $F_2$ at $y = 1$ ($x = Q^2/s$) in order to make its value equal to $2xF_1$. For historical consistency, we call this correction term "$F_L$", where $F_2 - F_L = 2xF_1$ even though the polarization of the virtual photon at this kinematic point is transverse. In this context, the $F_L$ measurements at $x_{min}$ favor the process $\gamma^* + (q_i, \overline{q_i}) \rightarrow g + (q_f, \overline{q_f})$ over gluon dominated processes. In fact, this observation is not surprising in that at y = 1 ($x_{min}$) the exchanged photon is transversely polarized and thus can couple with quark-antiquark pairs in a spin 1 state, some of which can be resonant in a virtual vector meson state, such as the ϕ meson. The chief conclusion of this paper can be succinctly expressed by the following limit of the reduced cross section as y → 1, x → $Q^2/s$:

$$\lim_{y\to 1}\left[\sigma_r\left(x,Q^2,s\right)\right]=2xF_1=F_2-F_L\approx F_2-E\frac{\alpha_s(Q^2)}{\pi}\left[\frac{4}{3}\int_x^1\frac{dy}{y}\left(\frac{x}{y}\right)^2F_2(y,Q^2)\right], \qquad (8)$$

where $F_L$ at $x_{min}$ is determined by $F_2$ enhanced by the factor E = 4.7 ± 0.5, a value that appears to be independent of √s over the HERA range 225 ≤ √s ≤ 318 GeV. The data suggest that the so called dipole contribution becomes important at the y = 1 limit of the reduced cross section. There are three observations of the structure functions $F_2$ and $F_L$ at $x_{min}$ that lead us to this conclusion:

1. The observation of the anomaly at Q ~ 1 GeV/c in $F_2$ of Table 1 is consistent with the strange quark-antiquark threshold, favoring the dipole picture (Fig. 6).
2. The behavior of the ratio of two statistically independent measurements, $(2xF_1)/F_2$ as a function of $Q^2$, strongly disfavors a large gluon component in $F_L$ (Fig. 8).
3. The behavior of $F_L$ as $Q^2$ goes to zero has a value that is significantly larger than would be expected if it were dominated by the gluon component. $F_L$ can be approximated by enhancing the $F_2$ component in Eq. 4 by a factor E = 4.7 ± 0.5, while turning off the gluon contribution completely (Fig. 10). It is a challenge to theory to estimate the value of E.

These observations do not resolve the discrepancies of the HERA data with the QCD SM analyses described by Abramowicz, et al. [3]. But they do suggest that the interaction of the virtual photon in e-p DIS at $x_{min} = Q^2/s$ involves more complicated physics than a simple QED coupling to a single quark charge. We attribute the puzzling turn-over at low x and $Q^2$ of the reduced cross section to our finding that $F_L$ is dominated by the $F_2$ component and is larger at low $Q^2$ than the gluon-dominated expectation. It therefore becomes a question whether or not the tension with the QCD analyses at very low $x \to x_{min} = Q^2/s$ lies in the structure of the nucleon itself or in the nature of virtual photon probe, or in both. It seems that the nature of the virtual photon probe at $x_{min}$ is a key factor.

For complete information, we tabulate our determinations of $F_2$, $2xF_1$ and $F_L$ in the Appendix.

**ACKNOWLEDGEMENTS**

The author thanks his MIT colleagues, especially Richard Milner, for stimulating discussions. And thanks to MIT Laboratory for Nuclear Science for the use of the subMIT computer facility. This note was written in the memory of Louis N. ('Lou') Hand.

# APPENDIX

Tabulated are the $F_2$, $2xF_1$ and $F_L$ values of this analysis for various values of √s as a function of $x_{min}= Q^2/s$ and $Q^2$ $(GeV/c)^2$. Only the statistical errors are shown. The systematic errors are estimated to be ± 10% for $F_2$ and $2xF_1$ and roughly ± 15% for $F_L$. Tables 1, 2, 3&4 are for $e^+p$ NC DIS and Table 5 is for $e^-p$ NC DIS. The raw data are from HEPData [4]. The $F_2$ values for HERA Table 2 are by a parameterization.

√s 318 GeV Table 1 $e^+p$

| $Q^2$ | $x_{min}$ | $F_2$ | ± | error | $2xF_1$ | ± | error | $F_L$ | ± | error |
|---|---|---|---|---|---|---|---|---|---|---|
| 0.20 | $1.98 \times 10^{-6}$ | 0.249 | ± | 0.008 | | | | | | |
| 0.25 | $2.47 \times 10^{-6}$ | 0.306 | ± | 0.007 | | | | | | |
| 0.35 | $3.47 \times 10^{-6}$ | 0.375 | ± | 0.012 | | | | | | |
| 0.40 | $3.95 \times 10^{-6}$ | 0.385 | ± | 0.009 | | | | | | |
| 0.50 | $4.94 \times 10^{-6}$ | 0.424 | ± | 0.017 | | | | | | |
| 0.65 | $6.42 \times 10^{-6}$ | 0.521 | ± | 0.030 | | | | | | |
| 0.85 | $8.41 \times 10^{-6}$ | 0.743 | ± | 0.043 | | | | | | |
| 1.20 | $1.19 \times 10^{-5}$ | 0.821 | ± | 0.035 | 0.441 | ± | 0.012 | 0.380 | ± | 0.037 |
| 1.50 | $1.48 \times 10^{-5}$ | 0.886 | ± | 0.025 | 0.502 | ± | 0.012 | 0.385 | ± | 0.028 |
| 2.00 | $1.98 \times 10^{-5}$ | 0.969 | ± | 0.024 | 0.656 | ± | 0.014 | 0.313 | ± | 0.028 |
| 2.70 | $2.67 \times 10^{-5}$ | 1.062 | ± | 0.023 | 0.740 | ± | 0.011 | 0.322 | ± | 0.026 |
| 3.50 | $3.46 \times 10^{-5}$ | 1.153 | ± | 0.021 | 0.771 | ± | 0.014 | 0.383 | ± | 0.025 |
| 4.50 | $4.45 \times 10^{-5}$ | 1.278 | ± | 0.026 | 0.896 | ± | 0.015 | 0.382 | ± | 0.030 |
| 6.50 | $6.43 \times 10^{-5}$ | 1.312 | ± | 0.024 | 1.011 | ± | 0.014 | 0.302 | ± | 0.028 |
| 8.50 | $8.40 \times 10^{-5}$ | 1.438 | ± | 0.027 | 1.109 | ± | 0.016 | 0.329 | ± | 0.031 |
| 10.0 | $9.89 \times 10^{-5}$ | 1.513 | ± | 0.052 | | | | | | |
| 12.0 | $1.19 \times 10^{-4}$ | 1.536 | ± | 0.030 | 1.248 | ± | 0.015 | 0.288 | ± | 0.034 |
| 15.0 | $1.48 \times 10^{-4}$ | 1.624 | ± | 0.025 | 1.276 | ± | 0.021 | 0.348 | ± | 0.033 |
| 18.0 | $1.78 \times 10^{-4}$ | 1.646 | ± | 0.032 | 1.325 | ± | 0.017 | 0.321 | ± | 0.036 |

| | | | | | | | | | | | |
|---|---|---|---|---|---|---|---|---|---|---|---|
| 22.0 | $2.18 \times 10^{-4}$ | 1.759 | ± | 0.041 | 1.367 | ± | 0.017 | 0.392 | ± | 0.044 |
| 27.0 | $2.67 \times 10^{-4}$ | 1.807 | ± | 0.036 | 1.430 | ± | 0.019 | 0.377 | ± | 0.040 |
| 35.0 | $3.46 \times 10^{-4}$ | 1.721 | ± | 0.025 | 1.395 | ± | 0.014 | 0.326 | ± | 0.028 |
| 45.0 | $4.45 \times 10^{-4}$ | 1.763 | ± | 0.030 | 1.450 | ± | 0.015 | 0.313 | ± | 0.034 |
| 60.0 | $5.93 \times 10^{-4}$ | 1.773 | ± | 0.024 | 1.487 | ± | 0.019 | 0.286 | ± | 0.031 |
| 70.0 | $6.92 \times 10^{-4}$ | 1.790 | ± | 0.044 | 1.490 | ± | 0.026 | 0.301 | ± | 0.051 |
| 90.0 | $8.90 \times 10^{-4}$ | 1.711 | ± | 0.045 | 1.489 | ± | 0.025 | 0.223 | ± | 0.051 |
| 120.0 | $1.19 \times 10^{-3}$ | 1.627 | ± | 0.053 | 1.410 | ± | 0.018 | 0.216 | ± | 0.056 |
| 150.0 | $1.48 \times 10^{-3}$ | 1.564 | ± | 0.048 | | | | | | |
| 200.0 | $1.98 \times 10^{-3}$ | 1.492 | ± | 0.039 | 1.305 | ± | 0.020 | 0.187 | ± | 0.044 |
| 250.0 | $2.47 \times 10^{-3}$ | 1.443 | ± | 0.051 | 1.269 | ± | 0.022 | 0.174 | ± | 0.055 |
| 300.0 | $2.97 \times 10^{-3}$ | 1.347 | ± | 0.049 | 1.229 | ± | 0.022 | 0.118 | ± | 0.054 |
| 400.0 | $3.96 \times 10^{-3}$ | 1.274 | ± | 0.056 | 1.287 | ± | 0.025 | -0.013 | ± | 0.061 |
| 500.0 | $4.94 \times 10^{-3}$ | 1.205 | ± | 0.071 | 1.119 | ± | 0.025 | 0.087 | ± | 0.075 |
| 650.0 | $6.43 \times 10^{-3}$ | 1.207 | ± | 0.094 | 1.021 | ± | 0.023 | 0.185 | ± | 0.097 |
| 800.0 | $7.91 \times 10^{-3}$ | 0.985 | ± | 0.073 | 0.998 | ± | 0.024 | -0.013 | ± | 0.077 |
| 1000.0 | $9.89 \times 10^{-3}$ | 0.907 | ± | 0.157 | 0.887 | ± | 0.028 | 0.020 | ± | 0.159 |
| 1200.0 | $1.19 \times 10^{-2}$ | 0.907 | ± | 0.120 | 0.885 | ± | 0.021 | 0.021 | ± | 0.122 |
| 1500.0 | $1.48 \times 10^{-2}$ | 0.794 | ± | 0.091 | | | | | | |
| 2000.0 | $1.98 \times 10^{-2}$ | 0.623 | ± | 0.194 | 0.750 | ± | 0.032 | -0.126 | ± | 0.197 |
| 3000.0 | $2.97 \times 10^{-2}$ | 0.513 | ± | 0.155 | 0.616 | ± | 0.033 | -0.103 | ± | 0.159 |
| 5000.0 | $4.94 \times 10^{-2}$ | 0.535 | ± | 0.185 | 0.493 | ± | 0.038 | 0.043 | ± | 0.189 |
| 8000.0 | $7.91 \times 10^{-2}$ | | | | 0.365 | ± | 0.038 | | | |
| 12000.0 | $1.19 \times 10^{-1}$ | | | | 0.197 | ± | 0.057 | | | |

| √s | 300 | GeV | | Table 2 | $e^+p$ | | | | | |
|---|---|---|---|---|---|---|---|---|---|---|
| $Q^2$ | $x_{min}$ | "$F_2$" | ± | e | $2xF_1$ | ± | e | $F_L$ | ± | e |
| 0.065 | $7.23 \times 10^{-7}$ | 0.129 | ± | 0.008 | 0.040 | ± | 0.002 | 0.089 | ± | 0.008 |
| 0.085 | $9.47 \times 10^{-7}$ | 0.149 | ± | 0.009 | 0.065 | ± | 0.005 | 0.084 | ± | 0.010 |
| 0.110 | $1.22 \times 10^{-6}$ | 0.173 | ± | 0.010 | 0.083 | ± | 0.005 | 0.090 | ± | 0.011 |
| 0.150 | $1.66 \times 10^{-6}$ | 0.210 | ± | 0.013 | 0.130 | ± | 0.005 | 0.081 | ± | 0.014 |
| 0.200 | $2.22 \times 10^{-6}$ | 0.255 | ± | 0.015 | 0.130 | ± | 0.007 | 0.125 | ± | 0.017 |
| 0.250 | $2.78 \times 10^{-6}$ | 0.296 | ± | 0.018 | 0.169 | ± | 0.010 | 0.127 | ± | 0.020 |
| 0.350 | $3.89 \times 10^{-6}$ | 0.370 | ± | 0.022 | 0.197 | ± | 0.012 | 0.172 | ± | 0.025 |
| 0.399 | $4.44 \times 10^{-6}$ | 0.402 | ± | 0.024 | 0.119 | ± | 0.045 | 0.283 | ± | 0.051 |
| 0.650 | $7.22 \times 10^{-6}$ | 0.542 | ± | 0.033 | 0.315 | ± | 0.057 | 0.228 | ± | 0.065 |
| 0.850 | $9.45 \times 10^{-6}$ | 0.632 | ± | 0.038 | 0.398 | ± | 0.102 | 0.234 | ± | 0.109 |
| 1.20 | $1.33 \times 10^{-5}$ | 0.756 | ± | 0.045 | 0.548 | ± | 0.100 | 0.208 | ± | 0.110 |
| 2.00 | $2.22 \times 10^{-5}$ | 0.952 | ± | 0.057 | 0.647 | ± | 0.066 | 0.305 | ± | 0.087 |
| 2.70 | $3.00 \times 10^{-5}$ | 1.069 | ± | 0.064 | 0.850 | ± | 0.040 | 0.219 | ± | 0.076 |
| 3.50 | $3.89 \times 10^{-5}$ | 1.168 | ± | 0.070 | 0.781 | ± | 0.082 | 0.387 | ± | 0.108 |
| 4.50 | $5.00 \times 10^{-5}$ | 1.260 | ± | 0.076 | 0.938 | ± | 0.061 | 0.322 | ± | 0.097 |
| 6.50 | $7.23 \times 10^{-5}$ | 1.384 | ± | 0.083 | 1.028 | ± | 0.035 | 0.356 | ± | 0.090 |
| 8.50 | $9.44 \times 10^{-5}$ | 1.465 | ± | 0.088 | 1.092 | ± | 0.099 | 0.372 | ± | 0.132 |
| 10.00 | $1.11 \times 10^{-4}$ | 1.510 | ± | 0.091 | 1.023 | ± | 0.085 | 0.486 | ± | 0.124 |
| 12.0 | $1.33 \times 10^{-4}$ | 1.556 | ± | 0.093 | 1.098 | ± | 0.048 | 0.458 | ± | 0.105 |
| 15.0 | $1.67 \times 10^{-4}$ | 1.605 | ± | 0.096 | 1.145 | ± | 0.050 | 0.460 | ± | 0.108 |
| 18.0 | $2.00 \times 10^{-4}$ | 1.641 | ± | 0.098 | 1.219 | ± | 0.099 | 0.422 | ± | 0.140 |
| 27.0 | $3.00 \times 10^{-4}$ | 1.702 | ± | 0.102 | 1.347 | ± | 0.032 | 0.355 | ± | 0.107 |
| 35.0 | $3.89 \times 10^{-4}$ | 1.727 | ± | 0.104 | 1.446 | ± | 0.094 | 0.281 | ± | 0.140 |
| 45.0 | $5.00 \times 10^{-4}$ | 1.741 | ± | 0.104 | 1.493 | ± | 0.085 | 0.247 | ± | 0.135 |
| 70.0 | $7.78 \times 10^{-4}$ | 1.737 | ± | 0.104 | 1.614 | ± | 0.093 | 0.123 | ± | 0.140 |
| 120.0 | $1.33 \times 10^{-3}$ | 1.676 | ± | 0.101 | 1.405 | ± | 0.092 | 0.272 | ± | 0.136 |
| 200.0 | $2.22 \times 10^{-3}$ | 1.550 | ± | 0.093 | 1.201 | ± | 0.104 | 0.349 | ± | 0.140 |
| 300.0 | $3.33 \times 10^{-3}$ | 1.394 | ± | 0.084 | 1.238 | ± | 0.097 | 0.156 | ± | 0.128 |
| 400.0 | $4.44 \times 10^{-3}$ | 1.289 | ± | 0.077 | 0.860 | ± | 0.086 | 0.429 | ± | 0.116 |
| 500.0 | $5.56 \times 10^{-3}$ | 1.212 | ± | 0.073 | 1.059 | ± | 0.125 | 0.154 | ± | 0.144 |
| 650.0 | $7.22 \times 10^{-3}$ | 1.123 | ± | 0.067 | 0.930 | ± | 0.049 | 0.193 | ± | 0.083 |
| 800.0 | $8.89 \times 10^{-3}$ | 1.054 | ± | 0.063 | 0.997 | ± | 0.089 | 0.057 | ± | 0.109 |
| 1200.0 | $1.33 \times 10^{-2}$ | 0.923 | ± | 0.055 | 0.880 | ± | 0.049 | 0.044 | ± | 0.074 |
| 1500.0 | $1.67 \times 10^{-2}$ | 0.855 | ± | 0.051 | 0.762 | ± | 0.076 | 0.093 | ± | 0.092 |
| 2000.0 | $2.22 \times 10^{-2}$ | 0.771 | ± | 0.046 | 0.632 | ± | 0.078 | 0.139 | ± | 0.090 |
| 3000.0 | $3.33 \times 10^{-2}$ | 0.661 | ± | 0.040 | 0.659 | ± | 0.079 | 0.002 | ± | 0.088 |
| 5000.0 | $5.56 \times 10^{-2}$ | 0.536 | ± | 0.032 | 0.431 | ± | 0.060 | 0.104 | ± | 0.068 |
| 8000.0 | $8.89 \times 10^{-2}$ | 0.429 | ± | 0.026 | 0.304 | ± | 0.114 | 0.126 | ± | 0.117 |
| 12000.0 | $1.33 \times 10^{-1}$ | 0.341 | ± | 0.020 | 0.277 | ± | 0.107 | 0.064 | ± | 0.109 |
| 20000.0 | $2.22 \times 10^{-1}$ | 0.226 | ± | 0.014 | 0.136 | ± | 0.053 | 0.091 | ± | 0.055 |

√s 238 GeV Table 3&4 $e^+p$

| $Q^2$ | $x_{min}$ | $F_2$ | ± | e | $2xF_1$ | ± | e | $F_L$ | ± | e |
|---|---|---|---|---|---|---|---|---|---|---|
| 2.5 | $4.41 \times 10^{-5}$ | | | | 0.865 | ± | 0.022 | | | |
| 3.5 | $6.18 \times 10^{-5}$ | | | | 0.798 | ± | 0.015 | | | |
| 5.0 | $8.83 \times 10^{-5}$ | | | | 0.884 | ± | 0.014 | | | |
| 6.5 | $1.15 \times 10^{-4}$ | | | | 0.968 | ± | 0.015 | | | |
| 8.5 | $1.50 \times 10^{-4}$ | 1.338 | ± | 0.032 | 1.028 | ± | 0.015 | 0.311 | ± | 0.055 |
| 12.0 | $2.12 \times 10^{-4}$ | 1.369 | ± | 0.032 | 1.134 | ± | 0.015 | 0.235 | ± | 0.058 |
| 15.0 | $2.65 \times 10^{-4}$ | 1.535 | ± | 0.035 | 1.139 | ± | 0.014 | 0.395 | ± | 0.063 |
| 20.0 | $3.53 \times 10^{-4}$ | 1.517 | ± | 0.042 | 1.167 | ± | 0.021 | 0.350 | ± | 0.074 |
| 25.0 | $4.41 \times 10^{-4}$ | 1.540 | ± | 0.031 | 1.235 | ± | 0.016 | 0.305 | ± | 0.054 |
| 35.0 | $6.18 \times 10^{-4}$ | 1.571 | ± | 0.031 | 1.303 | ± | 0.017 | 0.269 | ± | 0.054 |
| 45.0 | $7.94 \times 10^{-4}$ | 1.493 | ± | 0.030 | 1.275 | ± | 0.019 | 0.219 | ± | 0.053 |
| 60.0 | $1.06 \times 10^{-3}$ | 1.560 | ± | 0.034 | 1.251 | ± | 0.022 | 0.309 | ± | 0.060 |
| 90.0 | $1.59 \times 10^{-3}$ | 1.445 | ± | 0.035 | 1.292 | ± | 0.025 | 0.153 | ± | 0.062 |
| 120.0 | $2.12 \times 10^{-3}$ | 1.488 | ± | 0.049 | 1.208 | ± | 0.027 | 0.280 | ± | 0.086 |
| 150.0 | $2.65 \times 10^{-3}$ | 1.401 | ± | 0.070 | 1.216 | ± | 0.042 | 0.185 | ± | 0.121 |
| 200.0 | $3.53 \times 10^{-3}$ | 1.328 | ± | 0.080 | 1.166 | ± | 0.051 | 0.163 | ± | 0.135 |
| 250.0 | $4.41 \times 10^{-3}$ | 1.254 | ± | 0.077 | 1.042 | ± | 0.055 | 0.213 | ± | 0.134 |
| 300.0 | $5.30 \times 10^{-3}$ | 1.158 | ± | 0.097 | 0.902 | ± | 0.059 | 0.257 | ± | 0.169 |
| 400.0 | $7.06 \times 10^{-3}$ | 1.076 | ± | 0.144 | 0.927 | ± | 0.069 | 0.149 | ± | 0.238 |
| 500.0 | $8.83 \times 10^{-3}$ | 1.020 | ± | 0.120 | 0.919 | ± | 0.079 | 0.101 | ± | 0.211 |
| 650.0 | $1.15 \times 10^{-2}$ | 0.676 | ± | 0.126 | 0.873 | ± | 0.097 | -.197 | ± | 0.239 |
| 800.0 | $1.41 \times 10^{-2}$ | 0.685 | ± | 0.145 | 0.506 | ± | 0.084 | 0.179 | ± | 0.239 |

| √s | 318 | GeV | Table 5 $e^-p$ | | | | | | | |
|---|---|---|---|---|---|---|---|---|---|---|
| $Q^2$ | $x_{min}$ | $F_2$ | ± | e | $2xF_1$ | ± | e | $F_L$ | ± | e |
| 90.0 | $8.90 \times 10^{-4}$ | | | | 1.526 | ± | 0.033 | | | |
| 120.0 | $1.19 \times 10^{-3}$ | | | | 1.416 | ± | 0.025 | | | |
| 150.0 | $1.48 \times 10^{-3}$ | 1.634 | ± | 0.053 | | | | | | |
| 200.0 | $1.98 \times 10^{-3}$ | 1.578 | ± | 0.035 | 1.308 | ± | 0.023 | 0.270 | ± | 0.042 |
| 250.0 | $2.47 \times 10^{-3}$ | 1.535 | ± | 0.041 | 1.296 | ± | 0.027 | 0.239 | ± | 0.049 |
| 300.0 | $2.97 \times 10^{-3}$ | 1.440 | ± | 0.041 | 1.199 | ± | 0.026 | 0.240 | ± | 0.048 |
| 400.0 | $3.96 \times 10^{-3}$ | 1.343 | ± | 0.041 | 1.182 | ± | 0.027 | 0.161 | ± | 0.049 |
| 500.0 | $4.94 \times 10^{-3}$ | 1.277 | ± | 0.055 | 1.105 | ± | 0.029 | 0.173 | ± | 0.062 |
| 650.0 | $6.43 \times 10^{-3}$ | 1.219 | ± | 0.051 | 1.082 | ± | 0.027 | 0.137 | ± | 0.058 |
| 800.0 | $7.91 \times 10^{-3}$ | 1.081 | ± | 0.049 | 1.050 | ± | 0.028 | 0.030 | ± | 0.057 |
| 1000.0 | $9.89 \times 10^{-3}$ | 1.009 | ± | 0.095 | 0.931 | ± | 0.034 | 0.078 | ± | 0.101 |
| 1200.0 | $1.19 \times 10^{-2}$ | 0.983 | ± | 0.064 | 0.977 | ± | 0.023 | 0.006 | ± | 0.068 |
| 1500.0 | $1.48 \times 10^{-2}$ | 0.865 | ± | 0.055 | | | | | | |
| 2000.0 | $1.98 \times 10^{-2}$ | 0.889 | ± | 0.115 | 0.893 | ± | 0.038 | -0.003 | ± | 0.121 |
| 3000.0 | $2.97 \times 10^{-2}$ | 0.926 | ± | 0.123 | 0.755 | ± | 0.041 | 0.171 | ± | 0.130 |
| 5000.0 | $4.94 \times 10^{-2}$ | | | | 0.671 | ± | 0.051 | | | |
| 8000.0 | $7.91 \times 10^{-2}$ | | | | 0.690 | ± | 0.062 | | | |
| 12000.0 | $1.19 \times 10^{-1}$ | | | | 0.828 | ± | 0.156 | | | |